\documentclass[aps,prd,floatfix,nofootinbib,showpacs,superscriptaddress]{revtex4}
\usepackage{graphicx,amsmath,amssymb,bm}
\usepackage{color}
\usepackage[utf8]{inputenc}
\definecolor{purple}{rgb}{0.5,0,0.5}
\definecolor{blue}{rgb}{0.0,0,0.9}
\usepackage[colorlinks=true, pdfstartview=FitV, citecolor= purple, linkcolor = blue,urlcolor=blue]{hyperref}
\usepackage{amssymb}
\usepackage{graphics}
\usepackage{multirow}
\usepackage{epsfig}
\usepackage{rotating}
\usepackage{longtable} 
\usepackage{url}
\usepackage{pdflscape}
\usepackage{cancel}
\pdfoutput=1
\usepackage{csquotes}
\usepackage{placeins}
\usepackage{scalerel}
\usepackage{bbm}

\usepackage{array}
\usepackage{rotating}
\newcommand{\Y}{\mathcal{Y}}
\newcommand{\V}{S}
\newcommand{\A}{P}
\usepackage{comment}
\begin{document}
\title{A minimal axion model for mass matrices with five texture-zeros
}

\author{Yithsbey Giraldo}
 \email{yithsbey@gmail.com}
 \affiliation{Departamento de F\'\i sica, Universidad de Nari\~no, A.A. 1175,  San Juan de Pasto, Colombia}
\author{R. Martinez}
\email{remartinezm@unal.edu.co}
\affiliation{Departamento  de  Física,  Universidad  Nacional  de  Colombia, Ciudad Universitaria,  K.45 No.26-85, Bogotá D.C., Colombia} 
\author{Eduardo Rojas}
\email{eduro4000@gmail.com}
\affiliation{Departamento de F\'\i sica, Universidad de Nari\~no, A.A. 1175, San Juan de Pasto, Colombia}
\author{Juan C. Salazar}
\email{jusala@gmail.com}
\affiliation{Departamento de F\'\i sica, Universidad de Nari\~no, A.A. 1175, San Juan de Pasto, Colombia}

\begin{abstract}
A model  with fermion and scalar fields charged under a Peccei-Queen~(PQ) symmetry is proposed. 
The PQ charges are chosen in such a way that they can reproduce mass matrices with five texture zeros, {which can generate} the fermion masses, the CKM matrix, and the PMNS matrix of the Standard Model~(SM).
To obtain this result, at least 4~Higgs doublets are needed. As we will see in the manuscript this is a highly non-trivial result since the texture zeros of the mass matrices impose a large number of restrictions. 
This model shows a route to understand the different scales of the SM by  extending it with a  multi-Higgs sector and an additional  PQ symmetry. Since the PQ charges are not universal, the model presents flavor-changing neutral currents~(FCNC) at the tree level, a feature that constitutes the main source of restrictions on the parameter space.
We report the allowed regions by lepton decays and compare them with those coming from the semileptonic decays $K^{\pm}\longrightarrow \pi \bar{\nu}\nu$.   
We also show the excluded regions and the projected bounds of future experiments for the axion-photon coupling as a function of the axion mass and compare it with the parameter space of our model.  


\end{abstract}

\maketitle

\section{Introduction}
{The discovery of the Higgs with a mass of 125~GeV, by the ATLAS~\cite{Aad:2012tfa} and CMS~\cite{Chatrchyan:2012ufa} collaborations}, is very important because it {provides} experimental support for spontaneous symmetry breaking, which is the mechanism that explains the origin of the masses of fermions and gauge bosons. Additionally, it opens {up} the possibility of new physics in the scalar sector, such as the two Higgs doublet model~\cite{He:2008qm,Grzadkowski:2009iz,Logan:2010nw,Boucenna:2011hy,He:2011gc,Bai:2012nv,He:2013suk,Cai:2013zga,Wang:2014elb,Drozd:2014yla,Campbell:2015fra,vonBuddenbrock:2016rmr,Carena:2015moc}, models with additional singlet scalar fields~\cite{Branco:2011iw}, or scalar fields that could be candidates for Dark Matter~\cite{McDonald:1993ex,LopezHonorez:2006gr,Barbieri:2006dq,Honorez:2010re,Carena:2015moc}. On the other hand, in the Standard Model (SM)~\cite{Glashow:1961tr,Weinberg:1967tq,Salam:1968rm}, symmetry breaking generates {a coupling of the Higgs to fermions}, proportional to their masses, which is consistent with experimental data. However, there are several orders of magnitude between the fermion mass hierarchies that cannot be explained {within the context of the SM}. Six masses must be defined for the up and down quarks, three Cabibbo-Kobayashi-Maskawa~(CKM) mixing angles, and a complex phase {that involves} CP violation. On the other hand, in the lepton sector, there are three masses {for} charged leptons, two squared mass differences {for} neutrinos, three mixing angles, and a complex phase {that involves} CP violation in the lepton sector. In this case, it is necessary to determine the mass of the lightest neutrino and the character of neutrinos, {whether they are} Dirac or Majorana fermions.

In the Davis experiment~\cite{Davis:1968cp}, which was designed to detect solar neutrinos, a deficiency in the {solar neutrino flux was first observed.} According {to} the results of Bahcall, {only one-third} of solar neutrinos would reach the Earth~\cite{Bahcall:2004cc}.
Neutrino oscillation was first proposed by Pontecorvo~\cite{Pontecorvo:1957qd}, and the precise mechanism of solar neutrino oscillations was proposed by Mikheyev, Smirnov, and Wolfenstein{, involving a} resonant enhancement of neutrino oscillations due to matter effects~\cite{Wolfenstein:1977ue,Mikheyev:1985zog}.
These observations have been confirmed by many experiments from four different sources: solar neutrinos as in Homestake~\cite{Davis:1968cp}, SAGE\cite{Davis:1968cp}, GALLEX \& GNO\cite{Kaether:2010ag,Cleveland:1998nv}, SNO\cite{SNO:2011hxd}, Borexino\cite{Borexino:2008fkj,Bellini2014} and Super-Kamiokande\cite{Super-Kamiokande:2005wtt,Super-Kamiokande:2006jvq} experiments, atmospheric neutrinos as in IceCube\cite{IceCube:2014flw}, neutrinos from reactors as KamLAND\cite{KamLAND:2010fvi}, CHOOZ\cite{CHOOZ:1999hei}, Palo Verde\cite{Piepke:2002ju}, Daya Bay\cite{DayaBay:2015ayh}, RENO\cite{Kim:2016yvm} and SBL\cite{Kopp:2013vaa}, and from accelerators as in MINOS\cite{MINOS:2013utc}, T2K\cite{T2K:2013ppw}  and NO$\nu$A\cite{NOvA:2016kwd}.
Neutrino oscillations {depend} on squared mass differences{.} On the other hand, the lightest neutrino  mass {has not been determined yet}, but from cosmological considerations, none of the neutrino masses can exceed 0.3 eV, which implies that the neutrino masses are much smaller than the charged fermion masses.
However, unlike quarks and charged leptons,
in the SM the neutrinos are massless, which is explained by assuming that neutrinos are left-handed.
Therefore, the discovery of neutrino masses implies new physics beyond the SM. 
By adding right-handed neutrinos, the Higgs mechanism of the SM can give neutrinos the same type of mass {acquired by} charged leptons and quarks.
It is possible to add right-handed neutrinos~$\nu_R$ to the SM, as long as they do not participate in weak interactions.
With the presence of right-handed neutrinos, it would be possible to generate Dirac masses $m_{D}$, {similar} to those of charged leptons and quarks.
In principle, it is also possible to give Majorana masses to left-handed neutrinos, and similarly, right-handed neutrinos can have Majorana masses~$M_{R}$.
For a very large $M_{R}$, it would give effective Majorana masses for left-handed neutrinos as $m_{\text{eff}}\approx m_D^2/M_R$.
The presence of large Majorana masses {allows} to explain the tiny {neutrino masses} compared to the {charged fermion masses}~[38].
To explain the smallness of neutrino masses, there are three types of seesaw {mechanisms} in the literature: type I with three electroweak neutrinos and three heavy {right-handed} neutrinos, type II~\cite{Schechter:1980gr,Ma:1998dx}, type III~\cite{Foot:1988aq}, and inverse seesaw~\cite{Mohapatra:1986bd,Gonzalez-Garcia:1988okv}.
One way to explain the fermion mass {hierarchies} and the CKM and PMNS mixing angles is through zeros in the Yukawa couplings of fermions (this is known as texture-zeros or simply textures of the mass {matrices}, and these zeros are usually chosen by hand).
It is common in the literature to consider {Fritzsch-type textures~\cite{Branco:2007nb,Choubey:2008tb}, or similar~\cite{Adhikary:2011pv,Adhikary:2010fa,Branco:2005jr,Mei:2005qp,Hagedorn:2004ba,Antusch:2005gp}, for the neutrino and charged lepton} mass matrices.

There is no theory {that provides values for the entries of the Yukawa Lagrangian, and consequently, there is no a first-principle explanation for the masses and their large differences in the SM.} This lack of naturalness is known as the hierarchy problem~\cite{Ludl:2014axa,Ferreira:2014vna,Ludl:2015lta,Grimus:2004hf,Fonseca:2014koa}.
In this direction, a way that has been explored in the literature is to propose a sector with multiple scalar doublets {along with} discrete symmetries~{\cite{Giraldo:2020hwl,Bjorkeroth:2018ipq}}, to reduce the number of Yukawa couplings, or equivalently, {by introducing} texture-zeros in the mass matrices~\cite{Cheng:1987rs,Matsuda:2006xa,CarcamoHernandez:2006iuj,Langacker:2000ju,Leroux:2001fx,King:2005jy}.
It is also possible to consider global symmetry groups that prohibit certain Yukawas, {which} somehow generate the {texture-zeros} mentioned~\cite{Ludl:2014axa,Ferreira:2014vna,Ludl:2015lta,Grimus:2004hf,Fonseca:2014koa}.
Another way {of obtaining} these textures is through {horizontal gauge symmetries}, with the assignment of quantum numbers to the fermion sector, which can break the universality of the SM~\cite{Alvarado:2012xi,CarcamoHernandez:2006iuj,Ibanez:1994ig,Binetruy:1994ru,Nir:1995bu,Jain:1994hd,Dudas:1995yu,Pisano:1992bxx,Frampton:1992wt,Foot:1994ym,Long:1995ctv,Long:1996rfd,Long:1997vbr,Diaz:2002dx,Diaz:2003dk,Diaz:2004fs,Ochoa:2005ch}.
This gauge symmetry {generates} textures that {produce} flavor changes in the neutral currents and {that}, in principle, could be seen in future colliders.
There are models with electroweak extensions of the SM {such as} $SO(14)$, $SU(9)$, $3$-$3$-$1$, $U(1)_{X}$, etc.~\cite{Barr:1979xt,Taylor:1989tm,Barr:1989ta,Weinberg:1972ws,Mohapatra:1974wk,Barr:1976bk,Balakrishna:1987qd,Barr:2001vj,Ferretti:2006df,Barr:2008pn,Frampton:2009ce,Babu:1995hr,Sato:1997hv,Irges:1998ax,Barr:2000ka,Haba:2000be} {that attempt} to explain the flavor and the mass hierarchy problem of the SM.
{Another mechanism to generate textures in the Yukawa Lagrangian is through additional discrete or global symmetries.}
Some groups {that} have been used in the literature are: $S_3$, $A_4$, $\Delta_{27}$, $Z_2$, etc. \cite{He:2006dk,Ahn:2013mva,GonzalezFelipe:2013xok,Ishimori:2012fg,GonzalezCanales:2012blg,Mohapatra:2012tb,Ding:2013hpa,Ma:2013xqa,Nishi:2013jqa,Altarelli:2005yp,Ishimori:2010fs,Kadosh:2010rm,Ding:2013eca}.
The simplest symmetries are of abelian type, which can be used to impose texture-zeros {in} the mass matrices to make them predictive.
On the other hand, given fermion mass matrices with texture-zeros, it is possible to find an extended scalar sector so that the texture-zeros can be generated from abelian symmetries~{\cite{Giraldo:2020hwl,Bjorkeroth:2018ipq}}.

Due {to the fact that there are three up-type quarks and three down-type quarks}, the mass operators are $3\times 3$ complex matrices with 36 degrees of freedom. {If we consider these operators to be} Hermitian~\cite{Giraldo:2011ya,Branco:1999nb,Giraldo:2015cpp}, the number of free parameters reduces to~18, which cannot be fully determined from the 10 {available physical quantities}, {namely masses and mixing angles}~\cite{Ramond:1993kv}.
This {provides} freedom to reduce the number of {free parameters in the matrices} and {search} for {matrix} structures with zeros that {provide} eigenvalues and mixing angles consistent with the {masses and mixing matrices of the fermions}.
One way to find zeros in {the} mass matrices that is automatically consistent with {experimental data} is based on weak basis transformations~(WBT) {for} quarks and leptons~\cite{Fritzsch:1999ee, Branco:1988iq, Branco:1999nb, Giraldo:2015cpp}. 
Fritzsch proposed an ansatz with six zeros~\cite{He:1989eh,Fritzsch:1977vd,Fritzsch:1977za,Fritzsch:1977vd,Fritzsch:1979zq,Du:1992iy,Fritzsch:2002ga}, but the value {of} $|V_{ub}/V_{cb}|\approx 0.06$ {is too small} compared to the experimental value~$|V_{ub}/V_{cb}|_{exp}\approx 0.09 $~\cite{ParticleDataGroup:2022pth}.
For this reason, {the use of 4 and 5 zero-textures was proposed}~\cite{Fritzsch:2002ga,Xing:2003yj,Branco:1999nb,Zhou:2003ji,Giraldo:2011ya,Benavides:2020pjx,Giraldo:2015ffa}.
Reference~\cite{Giraldo:2011ya,Giraldo:2015cpp} {showed} that matrices with five zero-textures could {reproduce} the mass hierarchy and mixing angles of the CKM~matrix.

The strong CP problem arises from the fact that the QCD Lagrangian has a non-perturbative term ("$\theta$-term") that explicitly violates {CP} in strong interactions.
Peccei and Quinn proposed a solution to the strong CP problem~\cite{Peccei:1977hh,Peccei:1977ur},
{where it is assumed that the SM has an additional global chiral symmetry~$U(1)$, which is spontaneously broken at} a large energy scale~$f_a$.
{One} consequence of this breaking is the existence of a particle called {the} axion, which is the Goldstone boson of the broken~$U(1)_{PQ}$ symmetry~\cite{Weinberg:1977ma,Wilczek:1977pj}.
Due to {the fact that} the PQ symmetry is not exact at the quantum level, as a result of a chiral anomaly, the axion is massive and its  {mass~(see Appendix~\ref{sec:appendix-gamma}) is given} by:
\begin{equation}
m_a=\frac{f_\pi m_\pi}{f_a}\frac{\sqrt{z}}{1+z}\approx 6 \mu eV\left( \frac{10^{12}GeV}{f_a}\right),
\end{equation}
where $z=0.56$ is assumed for the {up and down quark mass ratio}, while $f_\pi\approx 92$~MeV and $m_\pi = 135$~MeV are the pion decay constant and mass, respectively.

The effective couplings of axions to ordinary particles are inversely proportional to~$f_a$, {and also depend on the model.}
{It was originally thought that the PQ symmetry breaking occurred at} the electroweak scale, {but experiments have ruled this out.}
The mass of the axion and its coupling to matter and radiation scale as $1/f_a$, {making its direct detection extremely difficult.}
The combined limits {from unsuccessful searches} in nuclear and particle physics experiments and {from stellar evolution imply} that $f_a\geq 3\times10^9$~GeV~\cite{Kim:1979if}.
Furthermore, {there is} an upper limit of $f_a\leq10^{12}$GeV {that comes from} cosmology, {since} light axions are produced {in} abundance during the QCD phase transition~\cite{Turner:1989vc,Raffelt:1990yz,Preskill:1982cy,Abbott:1982af,Dine:1982ah}.
Hence, these models are generically {referred to as} "invisible" axion models and {remain} phenomenologically viable.
There are two classes of invisible axion {models} in the literature: KSVZ (Kim, Shifman, Vainshtein, and Zakharov)~\cite{Kim:1979if, Shifman:1979if} and DFSZ (Dine, Fischler, Srednicki, and Zhitnitsky)~\cite{Dine:1981rt, Zhitnitsky:1980tq}.
The main difference between KSVZ{-type} and DFSZ{-type} axions is that the former do not couple to ordinary quarks and leptons at tree level, but {instead} require an exotic quark {that ensures a nonzero} QCD anomaly to {generate} CP violation.
Depending on the assumed value of $f_a$, the existence of axions could have interesting consequences in astrophysics and cosmology.
The emission of axions produced in stellar plasma {through} their coupling to photons, electrons, and nucleons would provide a new {mechanism for energy loss in stars.}
This could accelerate the evolutionary process of stars and, therefore, shorten their {lifespan}.
Axions can also exist as primordial cosmic relics produced copiously in early times and could be candidates for dark matter.
Of course, if axions are candidates for dark matter~\cite{Sikivie:2009qn,Sikivie:2006ni,Hannestad:2003ye,Hannestad:2005df,Melchiorri:2007cd,Hannestad:2007dd,Hannestad:2008js}, they should be very light, {with} $m_a \leq 10 \mu eV$.
However, data from numerous laboratory experiments and astrophysical observations, together with the cosmological requirement that the contribution to the mass density of the Universe from relic axions does not {saturate} the Universe, restricts the allowed values of the axion mass {to} a range of $10^{-5} eV < m_a < 10^{-2}$~eV.
One source of axions would be the Sun{, which,} coupled to two photons, could be produced through the Primakoff conversion of thermal photons in the electric and magnetic fields of the solar plasma.
The limits are {primarily} useful {for complementing the arguments of stellar energy loss~\cite{Viaux:2013lha}} and the {searches for solar axions} by CAST at CERN~\cite{CAST:2008ixs} and the Tokyo axion helioscope~\cite{Inoue:2008zp}.

The  {axion-photon coupling~(see Appendix~\ref{sec:appendix-gamma})} can be calculated in chiral perturbation theory as~\cite{Peccei:1977hh,Peccei:1977ur}.
\begin{equation}
g_{a\gamma}=-\frac{\alpha}{2\pi f_a}\left( \frac{E}{N} - \frac{2}{3}\frac{z+4}{z+1}\right).
\end{equation}
This coupling and the axion mass are related to each other through the relation~$E/N$, which {depends} on the model and can be tested {in experiments.}

The strongest limits on the axion-electron coupling are {derived} from observations of stars with a dense {core}, where bremsstrahlung is very effective.
These conditions are {realized} in White Dwarfs and Red Giant Stars, where the evolution of a White Dwarf is a {cooling process} by photon radiation and neutrino emission, with the possible addition of new energy loss channels such as axions.
{Current numerical analysis suggest {a} limit of $g_{ae} \leq 2.8\times 10^{-13}$~\cite{DiLuzio:2020wdo}.}
In particular, using data from the Sloan Digital Sky Survey~(SDSS) and SuperCOSMOS Sky Survey~(SCSS)~\cite{MillerBertolami:2014rka}, they showed that the {axion-electron} coupling is approximately $1.4 \times 10^{-13}$.
More recent analysis of the data in Ref.~\cite{MillerBertolami:2014rka} found $g_{ae} = 1.5^{+0.6}_{-0.9} \times 10^{-13}$ ($95\%$ CL)~\cite{{Giannotti:2017hny,10.1093/mnras/sty1162}}. 
The two groups {studying the} axion-electron coupling are M5~\cite{Viaux:2013lha} and M3~\cite{2018arXiv180210357S}.
 {Their combination yields} the limit $g_{ae} = 1.6^{+ 0.29}_{- 0.34} \times 10^{-13}$. For a recent and  {comprehensive} review of  {axion physics, see}~\cite{DiLuzio:2020wdo}.

This document is organized as follows: In Section 2, we {review} the textures for the quark and lepton mass matrices {that will} be used in this work. We {also} write the real parameters of these matrices {in terms} of the masses of the SM fermions and two free parameters. In Section 3, we present the particle content of our model and the {necessary} PQ charges to generate the {mass matrix textures} presented in Section 2. In Section 4, we adjust the Yukawa couplings to obtain the masses of the charged leptons and neutrinos. 
It is important to {note that we cannot use the VEVs to adjust the lepton masses, as these were already adjusted to reproduce the quark masses.} 
It is also important to note that by using a seesaw mechanism, we can avoid adjusting the Yukawas, however, that is not our purpose in the present work. In Section 5, {we show the Lagrangian of our model}. In Section 6, we present some constraints {in} the parameter space, {as well as projected constraints {for upcoming} experimental results, both for experiments under construction} and in the {data-taking} phase.

\section{The Five texture-zero mass matrices}
The reason for dealing with texture zeros in the Standard Model~(SM) and its extensions is to simplify as much as possible the number of free parameters that allow us to see relationships between masses and mixings present in these models. 
The Yukawa Lagrangian is responsible for giving mass to SM fermions after spontaneous symmetry breaking. A first simplification, without losing generality, is to consider that the fermion mass matrices are Hermitian, so the number of free parameters for each sector of quarks and leptons is {reduced} to 18, but there is still {an excess} of parameters to reproduce the experimental data {provided} in the literature. Due to the lack of a model to make predictions, discrete symmetries can be used to prohibit some components in the Yukawa matrix, generating the so-called texture zeros for the mass matrices. In many works, instead of proposing discrete symmetries, texture zeros are proposed as practical and direct alternatives. The advantage of this approach is that it is possible to choose each mass matrix in an optimal {way} for the analytical treatment of the problem, and {at the same time adjust} the mixing angles and the masses of the fermions.
\subsection{Quark Sector}
We should keep in mind that six-zero textures in the SM have already been discarded because their predictions are outside the experimental ranges allowed; but, five-zero textures for quark mass matrices is a viable possibility \cite{Verma:2017ppl,Xing:2019vks,Fritzsch:1999ee,Desai:2000bu,Ludl:2015lta,Ponce:2013nsa}. Specifically, we chose the following five-zero 
textures because they fit well with experimental quark masses and mixing 
parameters~\cite{Giraldo:2011ya, Giraldo:2015cpp, Giraldo:2018mqi}:
{
\begin{equation}
\label{5.1y}
\begin{split}
M^{U}&=
\begin{pmatrix}
 0&0&C_u\\
0&A_u &B_u\\
C_u^*&B_u^*&D_u
\end{pmatrix},
\\
M^{D}&=
\begin{pmatrix}
 0&C_d&0\\
C_d^*&0&B_d\\
0&B_d^*&A_d
\end{pmatrix}.
\end{split}
\end{equation}}
In addition, the phases in  $M^{D}$ can be removed by a weak basis 
transformation~(WBT)~\cite{Giraldo:2011ya,Branco:1988iq,Branco:1999nb}, so that they 
are absorbed by the off-diagonal terms in~$M^{U}$. In this way,   the mass 
matrices~\eqref{5.1y}  can be rewritten as:
{
\begin{equation}
\label{eq2}
\begin{split}
M^{U}&
=
\begin{pmatrix}
 0&0&|C_u|e^{i\phi_{C_u}}\\
0&A_u &|B_u|e^{i\phi_{B_u}}\\
|C_u|e^{-i\phi_{C_u}}&|B_u|e^{-i\phi_{B_u}}&D_u
\end{pmatrix},
\\
M^{D}&=
\begin{pmatrix}
 0&|C_d|&0\\
|C_d|&0&|B_d|\\
0&|B_d|&A_d
\end{pmatrix},
\end{split}
\end{equation}}
By applying the trace and the determinant to the mass matrices~\eqref{eq2}, 
before and after the diagonalization process, the free real parameters of 
$M^{U}$ and $M^{D}$  can be written in terms of their masses:
\begin{subequations}
\label{e3.4}
\begin{align}
\label{3.18}
 D_u&=m_u-m_c+m_t-A_u,\\
\label{34a}
|B_u|&=\sqrt{\frac{(A_u-m_u)(A_u+m_c)(m_t-A_u)}{A_u}},\\
\label{35a}
|C_u|&=\sqrt{\frac{m_u\,m_c\,m_t}{A_u}},
\\
 A_d&=m_d-m_s+m_b,\\
\label{34b}
|B_d|&=\sqrt{\frac{(m_b-m_s)(m_d+m_b)(m_s-m_d)}{m_d-m_s+m_b}},\\
\label{35b}
|C_d|&=\sqrt{\frac{m_d\,m_s\,m_b}{m_d-m_s+m_b}}.
\end{align}
\end{subequations}
{A} possibility that works very well is to consider the second generation of 
quark masses {to be} negative, i.e., with eigenvalues $-m_c$ and $-m_s$. 
And $A_u$ is a free parameter, {whose value}, determined by the quark 
mass hierarchy, must be in the following range:
\begin{equation}
\label{eq5}
 m_u\le A_u\le m_t.
\end{equation}
The exact analytical {procedure for diagonalizing} the mass matrices~\eqref{eq2} is {indicated}
in Appendix~\ref{sec:mat-diag}.

\subsection{Lepton sector}
In this work, we {will consider} Dirac neutrinos. This is achieved, in part, by extending the SM  with right-handed neutrinos. 
In this way, we can carry out a treatment similar to that of the quark sector, that is, the mass matrices of the lepton sector can be considered Hermitian and {the weak basis transformation}~(WBT) can be applied~\cite{Branco:1999nb,Giraldo:2011ya}. In the literature, work has been done considering various texture-zeros for the Dirac mass matrices of the lepton sector~\cite{Hagedorn:2005kz,Ahuja:2009jj,Liu:2012axa,Verma:2013cza,Ludl:2014axa,Verma:2014lpa,Fakay:2014nea,Cebola:2015dwa,Gautam:2015kya,Ahuja:2016san,Singh:2018lao,Ahuja:2018fmw,Barranco:2012ci}. In our treatment, we are going to consider the following five-zero texture model studied in the paper~\cite{Benavides:2020pjx},
which can  accurately reproduce the 
Pontecorvo-Maki-Nakagawa-Sakata~(PMNS) mixing matrix $V_{\text{PMNS}}$~(mixing angles and the CP violating phase),
the charged lepton masses, and the squared mass differences in the normal mass ordering.
\begin{equation}
\label{eq:texturalep}
\begin{split}
M^N&=
\begin{pmatrix}
 0&|C_\nu|e^{ic_\nu}&0\\
|C_\nu|e^{-ic_\nu}&E_\nu &|B_\nu|e^{ib_\nu}\\
0&|B_\nu|e^{-ib_\nu}&A_\nu
\end{pmatrix},
\\
M^E&=
\begin{pmatrix}
 0&|C_{\ell}|&0\\
|C_{\ell}|&0&|B_{\ell}|\\
0&|B_{\ell}|&A_{\ell}
\end{pmatrix}.
\end{split}
\end{equation}
Without loss of generality, by using a WBT, the phases of the charged lepton mass matrix, $M^E$, can be absorbed into the entries $C_\nu$ and $B_\nu$ of the neutrino mass matrix, $M^N$.
Similarly, as was done in the case of the quark sector, the parameters present in the mass matrices of the lepton sector~\eqref{eq:texturalep} can be expressed in terms of {the masses of the charged leptons}
 $m_e,m_\mu$ and $m_\tau$ and {the masses  of the neutrinos} $m_1,m_2$ and $m_3$, in the normal ordering~($m_1<m_2<m_3$):
 \begin{subequations}
\label{parametrostexlep}
\begin{align}
\label{Al}
 A_\ell&=m_e-m_\mu+m_\tau,\\
\label{Bl}
|B_\ell|&=\sqrt{\frac{(m_\tau-m_\mu)(m_e+m_\tau)(m_\mu-m_e)}{m_e-m_\mu+m_\tau}},\\
\label{Cl}
|C_\ell|&=\sqrt{\frac{m_e m_\mu m_\tau}{m_e-m_\mu+m_\tau}},
\\
\label{Dn}
 E_\nu&=m_1-m_2+m_3-A_\nu,\\
\label{Bn}
|B_\nu|&=\sqrt{\frac{(A_\nu-m_1)(A_\nu+m_2)(m_3-A_\nu)}{A_\nu}},\\
\label{Cn}
|C_\nu|&=\sqrt{\frac{m_1 m_2 m_3}{A_\nu}},
\end{align}
\end{subequations}
%

where the values of the masses and the parameter  $A_\nu$ are given in Table~\ref{tab:input}. Furthermore, for the adjustment of the mass matrices~\eqref{eq:texturalep} it is very convenient to assume that the eigenvalues associated with the masses of the second family, $-m_2 $ and 
$-m_\mu$, are negative quantities. The exact diagonalizing matrices of the mass matrices~\eqref{eq:texturalep} are shown in {appendix}~\ref{sec:mat-diag},  equations~\eqref{32} and~\eqref{32x}.
\section{PQ symmetry and the minimal particle content}
\subsection{Yukawa Lagrangian and the PQ symmetry}
%

The texture-zeros of the mass matrices defined in the equations \eqref{eq2} and~\eqref{eq:texturalep} can be generated by imposing a Peccei-Queen   symmetry $U(1)_{PQ}$ on the  Lagrangian model, Eq.~\eqref{eq6}~\cite{Bjorkeroth:2018ipq,Garnica:2019hvn,Ringwald:2015dsf}.
As will be explained below,  the minimal Lagrangian that allows us to implement this symmetry  is given by~\cite{Giraldo:2020hwl,Brivio:2017ije}
\begin{align}
\mathcal{L}_{\text{LO}}& \supset 
(D_\mu\Phi^{\alpha})^\dagger D^\mu\Phi^{\alpha}
+\sum_{\psi}i\bar{\psi}\gamma^{\mu}D_\mu \psi
+\sum_{i=1}^{2} (D_\mu S_i)^\dagger D^\mu S_i\notag\\
&
- \Bigg(
\bar{q}_{Li}y^{D\alpha}_{ij}      \Phi^{\alpha}d_{Rj}  +
\bar{q}_{Li}y^{U\alpha}_{ij}\tilde\Phi^{\alpha}u_{Rj} \notag\\
&+
\bar{\ell}_{Li}y^{E\alpha}_{ij}  \Phi^{\alpha}e_{Rj}+
\bar{\ell}_{Li}y^{N\alpha}_{ij}\tilde\Phi^{\alpha}\nu_{Rj}  +\text{h.c} \Bigg)\notag\\
+&(\lambda_Q\bar{Q}_R Q_LS_2+\text{h.c})-V(\Phi,S_1,S_2)\,.
\label{eq6}
\end{align}
As it was shown in reference~\cite{Giraldo:2020hwl}, at least four Higgs doublets are {required} to generate the   {quark mass textures, therefore} $\alpha=1,2,3,4$.
In  \eqref{eq6} $i,j$ are family indices~(there is an implicit sum over repeated indices). The superscripts  $U$, $D$, $E$, $N$
 refer to  up-type  quarks, down-type quarks, electron-like and neutrino-like fermions, respectively; and
 $D_\mu =\partial_\mu+i\Gamma_\mu$ is the covariant derivative in the SM.
{The scalar potential $V(\Phi,S_1,S_2)$ is shown in appendix A (for further details, see reference~\cite{Giraldo:2020hwl})}.
In Eq.~\eqref{eq6}  $\psi$  stands for the SM fermion fields plus the heavy quark $Q$~(see Tables~\ref{tab:pcontent1} and~\ref{tab:pcontent2}).
As it is shown in Table~\ref{tab:pcontent2} the PQ charges of the heavy quark can be chosen in such a way that only the  interaction with the scalar singlet $S_2$ is allowed. 
%
We assign $Q_{\text {PQ}}$ charges for the left-handed quark doublets~($q_L$): $x_{q_i}$, right-handed up-type quark singlets~($u_R $): $x_{u_i}$, right-handed down-type quark singlets~($ d_R $): $x_{d_i}$, left-handed lepton doublets~($\ell_L$): $x_{\ell_i}$,
 right-handed charged leptons~($e_R$): $x_{e_i}$ and right-handed {Dirac} neutrinos~($\nu_R$): $x_{\nu_i}$ for each family~($ i = 1,2,3 $). We follow a similar notation for {the} scalar doublets, $x_{\phi_{\alpha}}$~ ($\alpha=1,2,3,4$), and  {the} scalar singlets $x_{_{S_{1,2}}}$. 

In this work, the PQ charges assigned to the quark sector and the scalar sector, {as well as the VEVs} assigned to the scalar doublets, will be the same {as those} assigned in~\cite{Giraldo:2020hwl}~(Tables~\eqref{tab:pcontent1} and~\eqref{tab:pcontent2}), 
and we will adjust the PQ charges of the lepton sector to {reproduce the texture-zeros given in Eq.}~\eqref{eq:texturalep}. 
To forbid a given entry in the lepton mass matrices, the corresponding sum of {PQ charges} 
must be equal to zero, so that we can obtain texture-zeros by imposing the following conditions:
 \begingroup
\renewcommand*{\arraystretch}{1.5} 
\begin{align}
\label{13x}
& M^{N}=
\begin{pmatrix}
0&x&0\\
x&x&x\\
0&x&x
\end{pmatrix}
\longrightarrow 
\begin{pmatrix}
 S_{11}^{N\alpha}\neq 0 & S_{12}^{N\alpha}= 0 & S_{13}^{N\alpha}\neq 0\\
 S_{21}^{N\alpha}= 0 & S_{22}^{N\alpha}   = 0 & S_{23}^{N\alpha}= 0 \\
 S_{31}^{N\alpha}\neq 0 & S_{32}^{N\alpha}   = 0 & S_{33}^{N\alpha}= 0 
\end{pmatrix},
\\
&
\label{14x}
M^{E}=
\begin{pmatrix}
0&x&0\\
x&0&x\\
0&x&x
\end{pmatrix}
\longrightarrow 
\begin{pmatrix}
 S_{11}^{E\alpha}\neq 0 & S_{12}^{E\alpha}    =0 & S_{13}^{E\alpha}\neq 0\\
 S_{21}^{E\alpha}    =0 & S_{22}^{E\alpha}\neq 0 & S_{23}^{E\alpha}    =0 \\
 S_{31}^{E\alpha}\neq 0 & S_{32}^{E\alpha}    =0 & S_{33}^{E\alpha}    =0 
\end{pmatrix},
\end{align}
\endgroup
where $S_{ij}^{N\alpha}= (-x_{\ell_i}+x_{\nu_j}-x_{\phi_\alpha})$ and $S_{ij}^{E\alpha}= (-x_{\ell_i}+x_{e_j}+x_{\phi_\alpha}) $.

{Since} the PQ charges of the Higgs doublets~($\alpha=1,2,3,4$) {are already given}, the possible solutions of \eqref{13x} and \eqref{14x} are strongly constrained. Table~\ref{tab:pcontent1} {provides a solution for the PQ charges of the lepton sector.}
\begin{table}[h]
\begin{center}
{\begin{tabular}{|ccccc|c|c|c|c|}
\hline  
Particles & Spin &$SU(3)_C$ &$SU(2)_L$ &   $U(1)_Y$  &$U_{\text{PQ}}(i=1)$& $U_{\text{PQ}}(i=2)$& $U_{\text{PQ}}(i=3)$&    $Q_{\text{PQ}}$\\
\hline
$q_{Li}$  & 1/2  & 3        &  2  & 1/6 &  $-2 s_1 + 2 s_2 + \alpha_q$  & $-s_1 + s_2 + \alpha_q$   & $\alpha_q$   & $x_{q_i}$     \\  
$u_{Ri}$  & 1/2  & 3        &  1  & 2/3  &  $s_1 + \alpha_q$ &  $s_2 + \alpha_q$  &  $-s_1 + 2 s_2 + \alpha_q$  & $x_{u_i}$     \\
$d_{Ri}$  & 1/2  & 3        &  1  & $-1/3$ &  $2 s_1 - 3 s_2 + \alpha_q$ &  $s_1 - 2 s_2 + \alpha_q$  & $-s_2 + \alpha_q$   & $x_{d_i}$     \\
$\ell_{Li}$  & 1/2  & 1        &  2  &   $-1/2$    & $-2s_1+ 2s_2+\alpha_\ell$    & $-s_1+s_2+\alpha_\ell$     &$\alpha_\ell$      & $x_{\ell_i}$    \\  
$e_{Ri}$  & 1/2  & 1        &  1  &    $-1$   & $2s_1-3 s_2+\alpha_\ell$    & $s_1-2s_2+\alpha_\ell$     & $- s_2+\alpha_\ell$     & $x_{e_i}$    \\
$\nu_{Ri}$& 1/2  & 1        &  1  &  0     & $-4s_1+5s_2+\alpha_\ell$    & $-s_1+ 2s_2+\alpha_\ell$     &  $s_2+\alpha_\ell$    & $x_{\nu_i}$   \\
\hline
\end{tabular}
\caption{Particle content. The subindex  $i=1,2,3$ stand for the family number in the interaction basis. 
 Columns 6-8 are the Peccei-Quinn charges, $Q_{PQ}$, {for each family of quarks and leptons in the SM}.
$s_1, s_2$ and $\alpha$ are real parameters, with $s_1\ne s_2$.}
\label{tab:pcontent1}
}
\end{center}
\end{table}

\begin{widetext}
\begin{table}[h]
\begin{center}
\begin{tabular}{|ccccccc|}
\hline   
Particles & Spin &$SU(3)_C$ &$SU(2)_L$ &$U(1)_Y$&     $U_{\text{PQ}}$              &$Q_{\text{PQ}}$\\
\hline
$\Phi_{1}$  & 0  & 1        &  2       &   1/2  &     $s_1$         &$x_{\phi_1}$   \\  
$\Phi_{2}$  & 0  & 1        &  2       &   1/2  &      $s_2$        &$x_{\phi_2}$   \\  
$\Phi_{3}$  & 0  & 1        &  2       &   1/2  &       $-s_1 + 2 s_2$             &$x_{\phi_3}$   \\
$\Phi_{4}$  & 0  & 1        &  2       &   1/2  &      $-3 s_1 + 4 s_2$            &$x_{\phi_4}$   \\  
$Q_L$       & 1/2& 3        &  0       &    0   &      $x_{Q_L}$    &$x_{Q_L}$      \\ 
$Q_R$       & 1/2& 3        &  0       &    0   &     $x_{Q_R}$     &$x_{Q_R} $     \\  
$S_1$       & 0  & 1        &  1       &    0   &      $s_1-s_2$    &$x_{_{S_1}}$   \\ 
$S_2$       & 0  & 1        &  1       &    0   &   $x_{Q_R}-x_{Q_L}$& $x_{_{S_2}}$  \\ 
\hline
\end{tabular}
\caption{Beyond the SM fields and their respective PQ charges.
 The parameters $s_1, s_2$ are reals, with $s_1\ne s_2$  and $x_{Q_R}\ne x_{Q_L}$.}
\label{tab:pcontent2}
\end{center}
\end{table}
\end{widetext}
In our model we include two scalar singlets $S_1$ and $S_2$ that break the global symmetry $U (1)_{PQ}$.
The  QCD anomaly of the PQ charges is 
\begin{align}
\label{eq:parametrization2}
N= 2\sum_{i}^3x_{q_i}-\sum_{i}^3 x_{u_i}-\sum_{i}^3 x_{d_i}+A_Q\,, 
\end{align} 
where $A_{Q}=x_{QL}-x_{QR}$ is the contribution to the anomaly of the heavy quark $Q$, which is a singlet under the electroweak gauge group, with left (right) Peccei-Quinn charges $x_{QL,R}$, respectively.
We can write the charges as a function of $N$ (since $N$ must be different from zero), such that 
\begin{align}\label{eq:parametrization}
s_1= \frac{N}{9}\hat{s}_1,\hspace{1cm}s_2=\frac{N}{9}\left(\epsilon+\hat{s}_1 \right),\ \  \text{with}
\hspace{0.5cm }\epsilon=1-\frac{A_{Q}}{N}\,, 
\end{align}
where $\hat{s}_1$ and  $\epsilon$ are arbitrary real numbers. 
To solve the strong CP problem {with} $N\ne 0$  and simultaneously  generate the 
texture-zeros in the mass matrices, it is necessary to maintain $\epsilon=\frac{9 (s_1-s_2)}{N} \ne 0$. 
With these definitions for Flavor-Changing Neutral Currents~(FCNC) observables,  the 
relevant parameters are $\hat{s}_1$ and $\epsilon$. 
This parameterization is quite convenient (for those cases where the parameters $\alpha_q$ and $\alpha_{\ell}$ are not relevant) {because} by fixing $N$ and $f_a$, we can vary $\hat{s}_1$ and $\epsilon$ for a fixed $\Lambda_{\text{PQ}}=f_a N $ in such a way that the parameter space naturally reduces to two dimensions.
%
\section{Naturalness of Yukawa couplings}
\subsection{The {mass matrices} in the quark sector}
In {reference~\cite{Giraldo:2020hwl}}, it was {shown} that to generate five texture zeros in the quark mass matrices~\eqref{5.1y}, as a consequence of a PQ~symmetry, it is necessary to include at least four scalar doublets in the model. After spontaneous symmetry breaking, the quark sector mass matrices take {on} the following form:
\begin{align}\label{eq:yij}
M^{U}=\hat{v}_{\alpha} y^{U\alpha}_{ij}= 
\begin{pmatrix}
       0 & 0 & y^{U1}_{13}\hat{v}_1\\\vspace{0.2cm}
       0 & y^{U1}_{22}\hat{v}_1 &  y^{U2}_{23}\hat{v}_2\\
       y^{U1^*}_{13}\hat{v}_1 & y^{U2^*}_{23}\hat{v}_2 & y^{U3}_{33}\hat{v}_3
\end{pmatrix}
,
\hspace{0.5cm}
M^{D}=\hat{v}_{\alpha} y^{D\alpha}_{ij}=
\begin{pmatrix}
			      0                    & |y^{D4}_{12}|\hat{v}_4 & 
0\\
			      |y^{D4}_{12}|\hat{v}_4 & 0                    & 
|y^{D3}_{23}|\hat{v}_3\\
			      0                    & |y^{D3}_{23}|\hat{v}_3 & 
y^{D2}_{33}\hat{v}_2    
\end{pmatrix},
\end{align}
where the $\hat{v}_i$ {are defined in terms of the vacuum expectation values,} $\hat{v}_i= v_i/\sqrt{2}$.
In~\cite{Giraldo:2020hwl} it was shown that the five-texture zeros~\eqref{eq2} are flexible enough { to 
set the quark Yukawa couplings close to 1 for most of them} (except {for} $y^{U2}_{23}$, $y^{D3}_{23}$ and $y_{13}^{U1}$), in this way we obtain: 
\begin{align}
\label{eq28}
\hat{v}_1 =  1.71 \,\text{GeV},\hspace{0.5cm}
\hat{v}_2 =  2.91 \, \text{GeV}, \hspace{0.5cm}
\hat{v}_3 =  174.085 \, \text{GeV},\hspace{0.5 cm}
\hat{v}_4 =  13.3  \,\text{MeV}.
\end{align}
As we can see, the hermiticity of the mass matrices is not {fully} achieved, but it is good to impose it for several reasons: (i) In the SM and its extensions, in which the right chirality fields are singlets under $SU(2)$, the mass matrices can be assumed Hermitian without losing generality, (ii) the previous fact allows us to consider Hermitian mass matrices, even after imposing an additional $PQ$ symmetry in the model, (iii) we can implement the WBT method{~\cite{Giraldo:2011ya}}, and (iv) {there is} an extensive {literature} {on} physically viable Hermitian mass matrices. It is important to {noticing} that the mass matrices in Eq.~\eqref{eq:yij} are Hermitian.
%
\subsection{The mass matrices in the lepton sector}
We can obtain the lepton mass matrices by starting from the Yukawa Lagrangian~\eqref{eq6}, which is invariant under the Peccei-Quinn~$U(1)_{PQ}$ symmetry, and taking into account the Yukawa parameters and expectation values~\eqref{eq28}. After the spontaneous symmetry breaking, the mass matrices for neutral and charged leptons are given respectively by{~\cite{Benavides:2020pjx}}:
\begin{equation}
M^{N}=\hat{v}_{\alpha} y^{N\alpha}_{ij}= 
\begin{pmatrix}
       0 & y^{N 1}_{12}\hat{v}_1 & 0\\\vspace{0.2cm}
       y^{N 4 }_{21}\hat{v}_4 &  y^{N 2}_{22}\hat{v}_2 &  y^{N 1}_{23}\hat{v}_1\\
       0 &  y^{N 3}_{32}\hat{v}_3 & y^{N 2}_{33}\hat{v}_2
\end{pmatrix}
,
\hspace{0.5cm}
M^{E}=\hat{v}_{\alpha} y^{E\alpha}_{ij}=
\begin{pmatrix}
			      0                    & |y^{E4}_{12}|\hat{v}_4 & 0\\
			      |y^{E4}_{12}|\hat{v}_4 & 0                    & |y^{E3}_{23}|\hat{v}_3\\
			      0                    & |y^{E3}_{23}|\hat{v}_3 & y^{E2}_{33}\hat{v}_2    
\end{pmatrix}.
\label{eq29}
\end{equation}
As we previously mentioned, at least four Higgs doublets are needed
to obtain the five texture-zeros for the chosen quark mass matrices.
 Our goal in this work is to keep the same number of Higgs doublets and their respective $PQ$ charges to 
generate the mass matrices and texture zeros for the lepton sector, Eq.~\eqref{eq:texturalep}. 
To get an Hermitian mass matrix $M^N$, it is necessary to impose  $y_{21}^{N4}/y_{12}^{N1*} =\hat{v}_1/\hat{v}_4$ and $y_{32}^{N3}/y_{23}^{N1*}=\hat{v}_1/\hat{v}_3$, requiring that the 
diagonal elements be real, i.e., $y^{N 2}_{22}=y^{N 2*}_{22}$ and  $y^{N 2}_{33}=y^{N 2*}_{33}$. On the other hand,
to obtain a symmetric mass matrix, $M^E$, for the charged leptons, it is sufficient to assume that the Yukawa couplings are Hermitian.
Through these choices it is possible to avoid  additional Higgs doublets.  

{Based} on the results of Table~\ref{tab:input}, Appendix~\ref{sec:mat-diag}, and the relationships established in~\eqref{parametrostexlep}, we find the following values for the Yukawa {couplings} of the lepton sector:
\begin{equation}
\label{YukLepC}
\begin{split}
|y^{E4}_{12}|& =  0.569582,\hspace{1.5cm}
|y^{E3}_{23}| =  0.00248291,  \hspace{1.1cm}
y^{E2}_{33} =  0.574472,\\
|y^{N1}_{12}| &=  4.74362\times 10^{-6} ,\hspace{0.5cm}
|y^{N4}_{21}| = 0.000609894,  \hspace{0.9cm}
y^{N2}_{22} =  6.68808 \times 10^{-6}, \\
|y^{N1}_{23}| &=  0.0000159881,\hspace{0.8cm}
|y^{N3}_{32}| =  1.57047 \times 10^{-7},  \hspace{0.5cm}
y^{N2}_{33} = 8.65364 \times 10^{-6}.
\end{split}
\end{equation}
To reproduce the neutrino masses quoted in~\cite{Benavides:2020pjx}, in the SM is required a Yukawa coupling around $10^{-14}$. In our case, the smallest Yukawa coupling is $10^{-7}$, which {significantly} reduces the fine-tuning in comparison {to that given by the SM.}

\section{The Effective  Lagrangian}
\label{sec:effectiveL}
The strongest constraints {on} non-universal PQ charges come from the FCNC.
To determine these constraints, we {start} by writing the most general  {next-to-leading} order~(NLO) effective Lagrangian as~\cite{Georgi:1986df,Gavela:2019wzg}:
\begin{align}\label{eq:lagrangian}
\mathcal{L}_{\text{NLO}}
=c_{a\Phi^\alpha}O_{a\Phi^\alpha}+c_1\frac{\alpha_1}{8\pi}O_{B}+c_2\frac{\alpha_2}{8\pi}O_{W}
+c_3\frac{\alpha_3}{8\pi}O_{G},
\end{align}
 $c_{a\Phi^\alpha}$ and $c_{1,2,3}$ are Wilson coefficients; $\alpha _{1,2,3}= 
\frac{g_{1,2,3}^2}{4\pi}$,  where  $g_{1,2,3}$ are the coupling strengths of 
the electroweak interaction in the interaction basis; 
and the Wilson operators are: 
 \begin{align}
O_{a\Phi}=& i\frac{\partial^\mu a}{\Lambda_{\text{PQ}}}   \left((D_{\mu}\Phi^\alpha)^{\dagger} \Phi^\alpha
                                                        -\Phi^{\alpha\dagger}(D_{\mu}\Phi^\alpha)\right),\hspace{0.5cm}
   O_{B}=  -\frac{ a}{\Lambda_{\text{PQ}}}B_{\mu\nu}\tilde B^{\mu\nu},\notag\\
   O_{W}=& -\frac{ a}{\Lambda_{\text{PQ}}}W_{\mu\nu}^a\tilde W^{a\mu\nu},\hspace{3.5cm}
   O_{G}=  -\frac{ a}{\Lambda_{\text{PQ}}}G_{\mu\nu}^a\tilde G^{a\mu\nu}\,,
\end{align}
where  
$B$,  $W^a$  and $G^a$ correspond to the gauge fields associated with the SM gauge groups  $U(1)_Y$,  $SU(2)_L$  
and  $SU(3)_C$, respectively.  $a$ is the axion field which corresponds to the 
 CP odd component of $S_1$.
It is possible to redefine the fields  {by} multiplying by a phase
\begin{align}\label{eq:redefinitions}
\Phi^{\alpha}   &\longrightarrow e^{i\frac{x_{_{\Phi^{\alpha}}}}{\Lambda_{\text{PQ}}}a}\Phi^{\alpha},\notag\\
\psi_L &\longrightarrow e^{i\frac{x_{\psi_L}}{\Lambda_{\text{PQ}}}a}\psi_L,\notag\\
\psi_R &\longrightarrow e^{i\frac{x_{\psi_R}}{\Lambda_{\text{PQ}}}a}\psi_R,\notag\\
S_i &\longrightarrow e^{i\frac{x_{_{S_i}}}{\Lambda_{\text{PQ}}}a}S_i.
\end{align}
In this expression, $x_{\psi}$ corresponds to the PQ charges of the SM fermions, i.e.,   $\{x_{\psi_{L,R}}\}= \{x_{q_i}$, $x_{u_i}$, $x_{d_i},x_{l_i}$, $x_{e_i}$, $x_{\nu_i}\}$  and  $\{x_{\Phi^{\alpha}}\}$ are the PQ charges of  the Higgs doublets~$\{\Phi^{\alpha}\}$. 
Replacing these definitions in the kinetic terms of Eq.~\eqref{eq6}, we obtain new contributions to the effective Lagrangian Eq.~\eqref{eq:lagrangian} (the NLO contributions in the non-derivative terms cancel out). The leading order~(LO) terms in $\Lambda^{-1}_{\text{PQ}}$
can be written as~\cite{Georgi:1986df,Brivio:2017ije}:
%
%
\begin{align}
\mathcal{L}_{\text{NLO}}\longrightarrow \mathcal{L}_{\text{NLO}}+\Delta \mathcal{L}_{\text{NLO}}, 
\end{align}
where 
\begin{align}\label{eq:deltaL}
\Delta \mathcal{L}_{\text{NLO}} = \Delta\mathcal{L}_{K^\Phi}+\Delta\mathcal{L}_{K^\psi}+
\Delta\mathcal{L}_{K^S}+
\Delta\mathcal{L}(F_{\mu\nu}),
\end{align}
with 
\begin{align}\label{eq:nloL}
 \Delta\mathcal{L}_{K^{\Phi}}=& 
ix_{\Phi^{\alpha}}\frac{\partial^\mu a}{\Lambda_{\text{PQ}}}   \left[(D_{\mu}\Phi^\alpha)^{\dagger} \Phi^\alpha
                                                        -\Phi^{\alpha\dagger}(D_{\mu}\Phi^\alpha)\right],\notag\\
\Delta\mathcal{L}_{K^\psi}=&
\frac{\partial_\mu a}{2\Lambda_{\text{PQ}}}\sum_{\psi}(x_{\psi_L}-x_{\psi_R})\bar{\psi}\gamma^{\mu}\gamma^{5}\psi
                                     -(x_{\psi_L}+x_{\psi_R})\bar{\psi}\gamma^{\mu}\psi,\notag\\
\Delta\mathcal{L}_{K^S}=&  
ix_{_{S_i}}\frac{\partial^\mu a}{\Lambda_{\text{PQ}}}   \left[(D_{\mu}S_i)^{\dagger} S_i
-S^{\dagger}_i (D_{\mu}S_i)\right]
+\text{h.c}\,.
\end{align}
The field redefinitions~\eqref{eq:redefinitions}  {induce} a modification in the measure of the functional path integral  whose effects can be obtained from the divergence of the axial-vector current: $J^{PQ5}_{\mu}= \sum_{\psi}(x_{\psi_L}-x_{\psi_R})\bar{\psi}\gamma_\mu\gamma^5\psi$~\cite{Bauer:2017ris},
\begin{align}\label{eq:divj5}
\partial^{\mu}J^{PQ5}_{\mu}=&\sum_{\psi}2i m_{\psi}
(x_{\psi_L}-x_{\psi_R})
\bar{\psi}\gamma^5\psi
-\sum_{\psi}(x_{\psi_L}-x_{\psi_R})\frac{\alpha_1    Y^2(\psi)}{2\pi}B_{\mu\nu}\tilde B^{\mu\nu}\notag\\
-&\sum_{\text{ $ SU(2)_L$  doublets}}
\hspace{-0.5cm}
x_{\psi_L}
\frac{\alpha_2}{4\pi}W_{\mu\nu}^a\tilde W^{a\mu\nu}
 -\sum_{ \text{$SU(3)$ triplets}}
 \hspace{-0.5cm}
 (x_{\psi_L}-x_{\psi_R})
 \frac{\alpha_3}{4\pi}G_{\mu\nu}^a\tilde G^{a\mu\nu},
\end{align}
where  the hypercharge  is normalized by
$Q=T_{3L}+Y$.
 The relation~\eqref{eq:divj5} is an on-shell relation, which is consistent  
 with the momentum of an on-shell axion. 
{Substituting this result into} $\mathcal{L}_{K^{\psi}}= \frac{\partial^{\mu}a}{2\Lambda_{\text{PQ}}}J^{PQ5}_{\mu}= 
-\frac{a}{2\Lambda_{\text{PQ}}}\partial^{\mu}J^{PQ5}_{\mu}$
we obtain new contributions to the {leading-order}  Wilson coefficients~\cite{Salvio:2013iaa}
\begin{align}
c_1& \longrightarrow c_1-\frac{1}{3}\Sigma q+\frac{8}{3}\Sigma u+\frac{2}{3}\Sigma d
-\Sigma \ell+ 2\Sigma e,\notag\\
c_2& \longrightarrow c_2-3\Sigma q-\Sigma \ell,\notag\\
c_3& \longrightarrow c_3-2\Sigma q +\Sigma u+\Sigma d-A_{Q},
\end{align}
where $\Sigma q\equiv x_{q_1}+x_{q_2}+x_{q_3}$.
The corresponding NLO Lagrangian is
\begin{align}\label{eq:gauge}
 \Delta\mathcal{L}(F_{\mu\nu})=
&\frac{ a}{\Lambda_{\text{PQ}}}\frac{\alpha_1}{8\pi}B_{\mu\nu}\tilde B^{\mu\nu}
\left(\frac{1}{3}\Sigma q-\frac{8}{3}\Sigma u-\frac{2}{3}\Sigma d+\Sigma \ell-2\Sigma e\right)\notag\\
+&\frac{ a}{\Lambda_{\text{PQ}}}\frac{\alpha_2}{8\pi}W_{\mu\nu}^a\tilde W^{a\mu\nu}
\left(3\Sigma q+\Sigma \ell\right)\notag\\
+&\frac{ a}{\Lambda_{\text{PQ}}}\frac{\alpha_3}{8\pi}G_{\mu\nu}^a\tilde G^{a\mu\nu}
\left(2\Sigma q -\Sigma u-\Sigma 
d+A  _{Q}\right).
\end{align}
It is convenient to define $c_3^{\text{eff}}=  c_3-2\Sigma q +\Sigma u+\Sigma d-A_{Q}=-N$.  
In our case,  $c_i=0$ and the {only} contributions to $c_{i}^{\text{eff}}$ come from the anomaly.
It is {customary}  to define  $\Lambda_{\text{PQ}}  = f_a \lvert c_3^{\text{eff}}\rvert $
to include the factor $c_3^{\text{eff}}$ in the normalization of the PQ charges.
{From now on,} we {will} assume that all the PQ charges are normalized in this way, so that $x_{\psi}$ {corresponds to}  $ x_{\psi}/c_3^{\text{eff}}$.
For normalized charges, $c_{3}^{\text{eff}}=1$,  { therefore}, we still  {maintain} the general form despite writing  all the expressions in terms of the effective scale~$f_a$.

The scalar fields and their PQ charges  are {the same as in} the reference~\cite{Giraldo:2020hwl}, so the scalar potential~$V(\Phi, S) $ is identical to that of the mentioned reference.
{With the VEVs and couplings} given  in~\cite{Giraldo:2020hwl}, the model reproduces {the mass of the SM Higgs}, {while} the masses of the exotic scalars are above the TeV~scale. 
This potential has the appropriate number of Goldstone bosons to give masses to the SM gauge bosons $Z^0, W^\pm$ and has an extra field that can be identified with the axion $a$.

\section{Low energy  constraints}
\subsection{Flavor changing neutral currents}
Due to {the non-universal PQ charges in our model}, a tree-level analysis of {flavor-changing neutral currents is necessary}.
 {As mentioned in reference~\cite{DiLuzio:2020wdo}, the strongest limits on the
 axion-quark  FCNC couplings come from meson decays in light mesons and missing energy.}
%
The {decays $K^{\pm}\rightarrow \pi^{\pm}a$}  provide the tightest limits (NA62 Collaboration \cite{NA62:2021zjw}) for the axion mass~\cite{DiLuzio:2020wdo}. 
Currently the most restrictive limits come from the {semileptonic} decays of  kaons $K^{\pm}\rightarrow\pi^{\pm}\bar{\nu}\nu$ and leptons $\ell_1\rightarrow \ell_2$+missing energy.
From the term $\Delta\mathcal{L}_{K^{\psi}}$, we obtain the vector and  axial couplings
for a {multi-Higgs sector model}, as shown in references~\cite{DiLuzio:2020wdo,Giraldo:2020hwl}  
\begin{align}
\Delta\mathcal{L}_{K^{\psi}}= -\partial_\mu a \bar{f}_i\gamma^{\mu}
\left(g_{af_if_j}^{V}+\gamma^5g_{af_if_j}^{A}\right)f_j,
\end{align}
where
\begin{align}\label{eq:av-couplings}
g_{af_if_j}^{V,A}= 
\frac{1}{2f_a c^{\text{eff}}_3}
\Delta^{Fij}_{V,A}, 
\end{align}
where  $\Delta^{Fij}_{V,A}= \Delta^{Fij}_{RR}(d)\pm \Delta^{Fij}_{LL}(q)$   
with $\Delta^{Fij}_{LL}(q)= \left(U^F_{L}x_{q}~U_L^{F\dagger}\right)^{ij}$  and  $\Delta^{Fij}_{RR}(d)= \left(U^F_{R}x_{d}~U_R^{F\dagger}\right)^{ij}$. In these expressions, $F$  stands for $U, D,N$  or $E$
and the $U_{L,R}^{F}$ are de diagonalizing matrices (see Appendix~\ref{sec:mat-diag}).
In Eq.~\eqref{eq:av-couplings}, we {normalize} the charges {using} $c^{\text{eff}}_3$, as explained in the last paragraph of section~\ref{sec:effectiveL}
(in other references $|c^{\text{eff}}_3|=|N|$ {is considered, corresponding} to the $SU(3)\times U(1)_{PQ}$ anomaly).
{The branching ratio for lepton decays} $\ell_i \rightarrow \ell_j \, a $  is given by~\cite{Bjorkeroth:2018dzu}
\[
\text{Br}(\ell_1\rightarrow\ell_2\,a)=\frac{m_{\ell_1}^3}{16\pi\Gamma(\ell_1)}\left(1-\frac{m_{\ell_2}^2}{m_{\ell_1}^2}\right)^3|g_{a\ell_1\ell_2}|^2,
\]
in this expression, {the} vector and axial couplings contribute 
in the same way
\[
|g_{a\ell_1\ell_2}|^2=|g_{a\ell_1\ell_2}^V|^2+|g_{a\ell_1\ell_2}^A|^2.
\]
where $m$ is the mass of the leptons and $\Gamma (\ell_i)$ is the total decay width of the particle $\ell_j$.

For the lepton decay  $\ell_i \rightarrow \ell_j \, a \, \gamma$, we can relate {this} branching {ratio} to the branching {ratio} of the process without the photon in the final state, according to the expression:
\[
\text{Br}(\ell_1\rightarrow\ell_2\,a\,\gamma)=\left(\frac{\alpha}{2\pi}\int dx \,dy f(x,y)\right)\text{Br}(\ell_1\rightarrow\ell_2\,a),
\]
where $\alpha$ is the fine structure constant,  and the function, $f(x,y)=\frac{(1-x)(2-y-xy)}{y^2(x+y-1)}$, depends on the mass and the energies  $x=2E_{\ell_2}/m_{\ell_1}$ and $y=2E_{\gamma}/m_{\ell_1}$.
For the lepton decay 
 $\mu\rightarrow e\,a\,\gamma$, the constraints come from the  Crystal Box experiment~\cite{Bolton:1988af}, with cut energies $E_\gamma,E_e>30\,\text{MeV}$, $\theta_{e\gamma}>140^\circ$, where $\text{cos}\,\theta_{e\gamma}=1+\frac{2(1-x-y)}{xy}$, so that $\int dx \,dy f(x,y)\approx0.011$.  
\begin{table}\label{tab:fcnc}
 \begingroup
\renewcommand*{\arraystretch}{1.5} 
 \begin{equation}
 \begin{array}{|l|l|}
 \hline 
 \hspace{0.5cm} \text{Collaboration} & \hspace{1cm}\text{Upper bound} \\
 \hline
 \text{N62 Collaboration\cite{NA62:2021zjw}} 
            &\mathcal{B}\left(K^+\rightarrow\pi^+      a\right)<(10.6^{+4.0}_{-3.4}\big{|}_{stat}\pm 0.9_{syst})\times 10^{-11}\\ 
\text{TRIUMF~\cite{Jodidio:1986mz} }
            &\mathcal{B}\left(\mu^+\rightarrow e^+ a   \right)<2.6\times 10^{-6} \\
\text{Crystal Box~\cite{TWIST:2014ymv} }
            &\mathcal{B}\left(\mu^+\rightarrow e^+\gamma a   \right)<1.1\times 10^{-9} \\
\text{ARGUS~\cite{Albrecht:1995ht} }
            &\mathcal{B}\left(\tau^+\rightarrow e^+ a   \right)<1.5\times 10^{-2} \\
\text{ARGUS~\cite{Albrecht:1995ht} }
            &\mathcal{B}\left(\tau^+\rightarrow \mu^+ a   \right)<2.6\times 10^{-2} \\
\hline
 \end{array}\notag
\end{equation}
  \caption{These inequalities come from the window for new physics in the branching ratio uncertainty of the meson decay in a pair $\bar{\nu}\nu$.}
 \endgroup  
\end{table}
\begin{figure}[h!]
\begin{center}
\centering 
\begin{tabular}{cc}
 \includegraphics[scale=0.5 ]{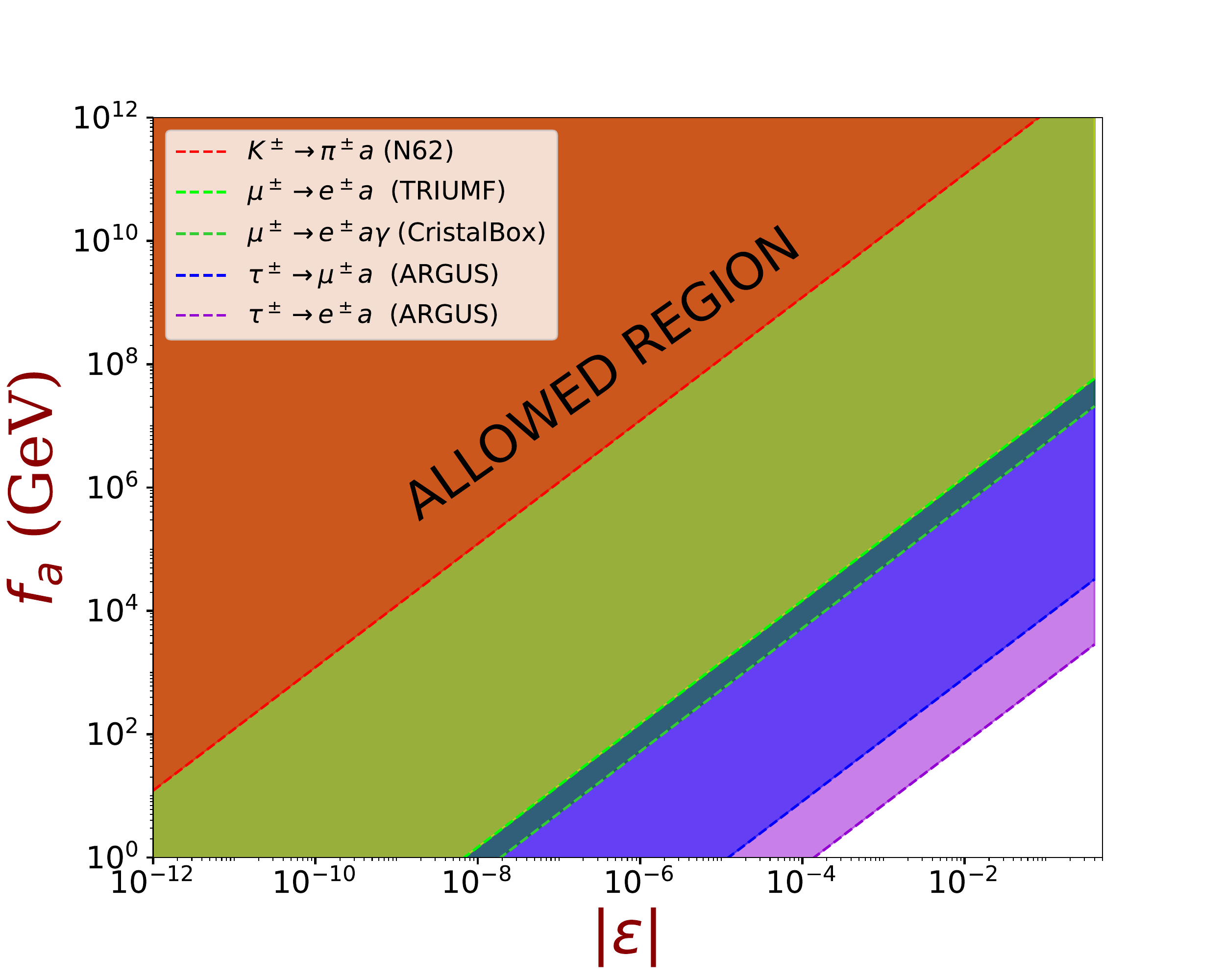}  
\end{tabular}
\end{center}
\caption{Allowed regions by lepton  decays.}
\label{fig:fa_vs_epsilon}	
\end{figure}
{
In our model, there is a natural alignment between the $\Phi_{3}$ (which is quite similar to $H_{1}$ in the Georgi basis~\cite{Georgi:1978ri}) and the standard model Higgs boson as a consequence of the large suppression of the VEVs of the scalar doublets $v_i$, with $i=1,2,4$, respect to $v_3$, the VEV of $\Phi_3$.
To some extent, this alignment avoids FCNC involving the SM Higgs boson~\cite{Georgi:1978ri}; however, after alignment, there are other sources of FCNC associated with the additional scalar doublets, which cannot be avoided by any means; 
however, as argued in Ref.~\cite{Giraldo:2020hwl} they are suppressed by a factor $1/M^4$ (where $M> 1$TeV is the mass of the exotic scalar doublets),
and therefore, our model avoids these potential sources of FCNC in agreement with  the general argument presented in~\cite{Georgi:1978ri}.} 

{From astrophysical considerations we have:  bounds from black holes superradiance and the SN 1987A upper limit on the neutron electric dipole moment, which, when combined, impose a constraint on the axion decay constant  in the range~\cite{DiLuzio:2020wdo} (see Figure~\ref{fig:fa_vs_epsilon}) :
$
    0.8\times10^{6}\text{GeV}\leq f_a\leq 2.8\times 10^{17}\text{GeV}
$.}

\subsection{Constraints on the axion-photon coupling}
\begin{figure}[h]
\begin{center}
\centering 
\includegraphics[scale=0.6]{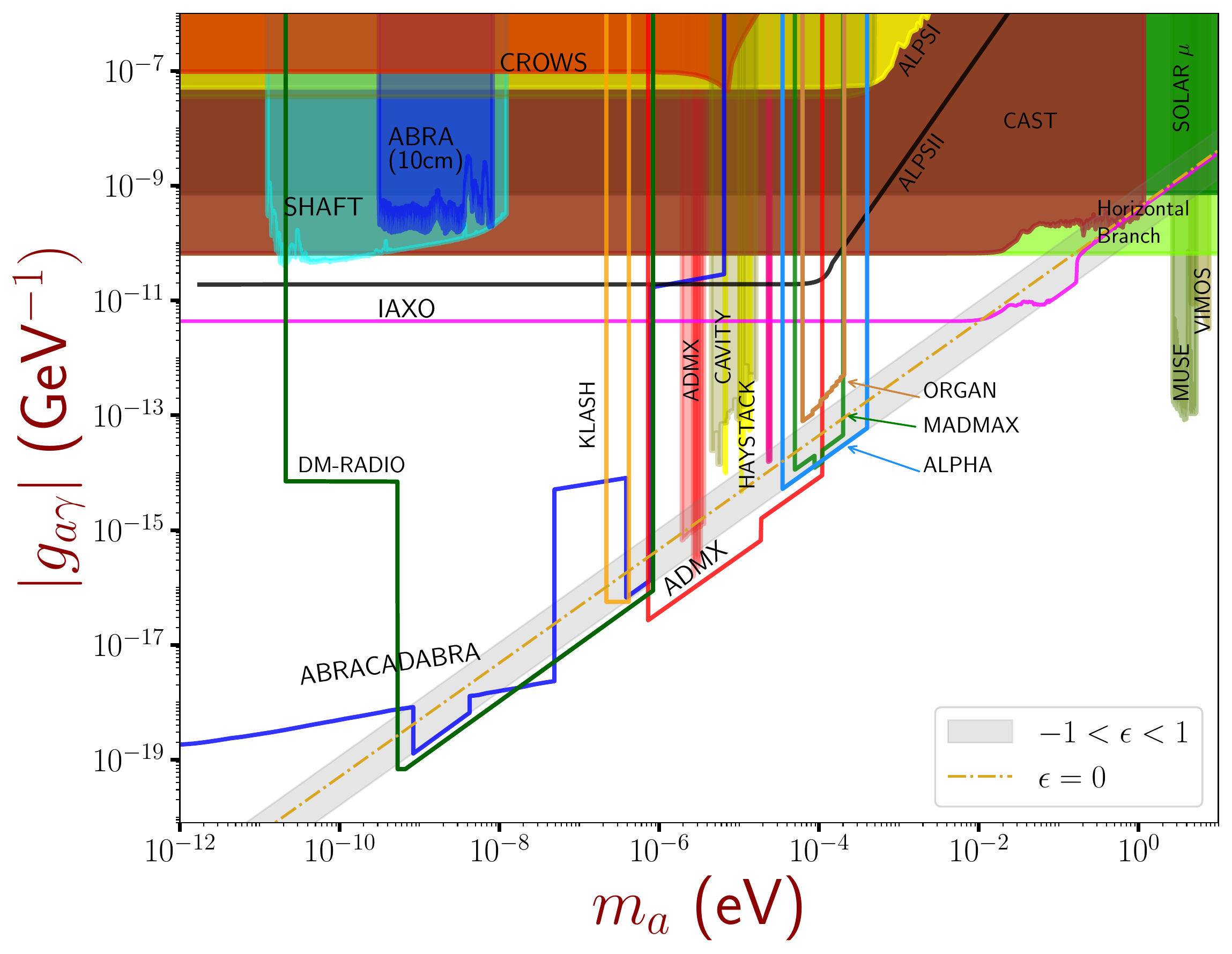}  
\end{center}
\caption{
The excluded parameter space by various experiments corresponds to the colored regions,
the dashed-lines correspond to the projected bounds of coming experiments looking for axion signals. The gray region corresponds to the parameter space scanned by our model.
}
\label{fig:ExperimentosAxion}	
\end{figure}

There are several experiments designed to look for  exotic particles. 
The sources {studied} in the search for axions are: the solar axion flux (helioscopes experiments), dark matter halo (haloscopes experiments), and axions produced in the laboratory.   

Among the experiments with the potential to search for evidence of axions in regions that {cover areas} within
the limits established by the parameters of our model are: DM-Radio \cite{Chaudhuri:2014dla}, KLASH \cite{Alesini:2017ifp, Gatti:2018ojx}, ADMX \cite{Stern:2016bbw}, ALPHA \cite{Lawson:2019brd}, MADMAX \cite{Beurthey:2020yuq}, IAXO \cite{Armengaud:2014gea, Shilon:2013bla}   and ABRACADABRA \cite{Ouellet:2018beu}. Similarly, some experiments have already ruled out regions established by the parameters of our model, among which are: ADMX \cite{ADMX:2021nhd, ADMX:2019uok,ADMX:2020hay}, CAST \cite{CAST:2017uph,CAST:2007jps},
CAPP\cite{Lee:2020cfj,Jeong:2020cwz,CAPP:2020utb}, HAYSTACK \cite{HAYSTAC:2020kwv,HAYSTAC:2018rwy}, Solar $\nu$ \cite{Gondolo:2008dd}, Horizontal Branch \cite{Ayala:2014pea}, MUSE \cite{Regis:2020fhw} and VIMOS \cite{Grin:2006aw} 

\section{Summary and conclusions }
We have presented a model {in which} the fermion and scalar fields are charged under a  $U(1)_{PQ}$ Peccei-Quinn symmetry. A recent work~\cite{Giraldo:2020hwl} showed that at least four Higgs doublets are required to generate  Hermitian mass matrices in the quark sector with five texture-zeros, {reproducing} the quark masses, the mixing angles, and the CP-violating phase of the CKM mixing matrix. In this work, we show that using the same number of  Higgs doublets, without changing the PQ charges in the quark and Higgs sectors, it is possible to generate  Hermitian mass matrices in the lepton sector {that} reproduce {the neutrino mass-squared differences} in the normal mass ordering,    the mixing angles, and the CP-violating phase of the PMNS mixing matrix. This result is quite non-trivial {as we maintain} the same four Higgs doublets required in the quark sector to generate a different texture pattern in the lepton sector.
 When {compared to} the SM, our model has almost all  Yukawa couplings close to 1 in the quark sector.  In the neutrino sector, the smallest Yukawa coupling is of the order of $1.6 \times 10^{-7}$, which is seven orders of magnitude larger than the corresponding Yukawa coupling in the SM, so {it requires} less fine-tuning than the SM.  

The polar decomposition theorem~\cite{VPrasolov,Hassani} allows any matrix to be written as the product of a Hermitian {matrix}  and a unitary matrix. In the SM and in theories where the right-handed fermion fields are singlets under the gauge group, it is possible to absorb the unitary matrix {into} the right-handed fields by redefining them; from this procedure, we can write any mass matrix as a Hermitian matrix.
In our work, we assume that the mass matrices are Hermitian in the interaction space,  this hypothesis has been used in {previous} studies on textures~\cite{Fritzsch:1999ee,Xing:2015sva,Gupta:2013yha,Ludl:2015lta,Giraldo:2011ya,Fusaoka:1998vc}, and {it is} quite useful for studying the flavor problem. In our work, we have normalized the PQ charges with the  QCD anomaly $ -N $ in such a way that {by} keeping the parameter $\epsilon \neq 0$,  we obtain  the textures of the mass matrices, 
addressing the flavor and strong CP problems simultaneously.

If nature is not fine-tuned in a more fundamental high-energy theory,
we {expect} that, eventually, it will be possible to find a texture that allows us to obtain all the scales of the SM from the VEVs of a Higgs sector with a {minimal scalar} content without the need to adjust the Yukawa couplings.

In our analysis, we report the constraints from lepton decays and compare them with the constraints from the search for neutrino pairs in charged Kaon decays $ K ^{\pm} \longrightarrow \pi^{\pm}\bar{\nu}\nu $. The results are shown in figure~\ref{fig:fa_vs_epsilon},
where the allowed region in the parameter space generated by $\epsilon $ and the axion decay constant $f_a$ {is displayed}.
This figure shows that the strongest {constraints} come from the semileptonic meson decay $K^{\pm}\rightarrow\pi \nu\bar{\nu}$.
It is important to note that the lepton decays do not {further constrain} the parameter space of our model ({compared to the region excluded} by the meson decay). 
We also show the excluded regions for the axion-photon coupling as a function of the axion mass; these results are summarized in Figure~\ref{fig:ExperimentosAxion}; the gray region corresponds to the parameter space of our model in the interval~$-1<\epsilon<1 $. 

In this article, we have {demonstrated} that with four Higgs doublets, it is possible to {fit} the textures of the mass matrices, both in the lepton and quark sectors. These matrices generate the masses and the mixing matrices for quarks and leptons within the {experimentally reported} values in the literature. {The introduction} these doublets improves the fine-tuning problem of {the} Yukawa couplings and {shows} that this approach is a {viable} way to {tackle} the flavor problem. We hope {to improve our results in future work by} using the Seesaw mechanism in the lepton sector.

\section*{Acknowledgments} 
We thank Financial support from ``Patrimonio Aut\'onomo Fondo Nacional de Financiamiento para la Ciencia, la Tecnolog\'ia
y la Innovaci\'on, Francisco Jos\'e de Caldas''.
This research was partly supported by the ``Vicerrectoría de Investigaciones e Interacción Social VIIS de la Universidad de Nariño'',  project numbers 1928, 2172,  2686,  2693 and 2679.
E.R and Y.G.   acknowledge additional financial support from Minciencias CD 82315 CT ICETEX 2021-1080.

\appendix
\section{The mass operator matrices} 
\label{sec:mass-op}
The most general Yukawa Lagrangian for the interaction of four Higgs  doublets $\Phi_\alpha$  with the SM fermions is given by
\begin{align}\label{eq:yukawa}
\mathcal{L}= 
 -\bar{q}_L^{\prime i} \Phi_\alpha y_{ij}^{D\alpha}d^{\prime j}_{R}
 -\bar{q}_L^{\prime i} \tilde\Phi_\alpha y_{ij}^{U\alpha}u^{\prime j}_{R}
 -\bar{\ell}_L^{\prime i} \Phi_\alpha y_{ij}^{E\alpha}e^{\prime j}_{R}
 -\bar{\ell}_L^{\prime i} \tilde\Phi_\alpha y_{ij}^{N\alpha}\nu^{\prime j}_{R} 
 +\text{h.c},
\end{align}
where a sum is assumed on repeated indices.
Here  $i,j$ run over $1,2,3$ and $\alpha$ over $1,2,3,4$. 
The Higgs  boson doublet fields are parameterized as follows:
{
\begin{align}
\Phi_\alpha = 
\begin{pmatrix}
\phi_\alpha^{+}\\
\frac{v_\alpha+h_\alpha+i\eta_\alpha}{\sqrt{2}}
\end{pmatrix},
\hspace{1cm}
\tilde\Phi_\alpha=i\sigma_2 \Phi_\alpha^{*}.
\end{align}}
{Similar to} the two Higgs doublet  model~\cite{Cardozo:2020uol} 
we  rotate the Higgs fields  to the (generalized) Georgi basis, {that is},
\begin{align}\label{eq:rab}
\begin{pmatrix}
H_1 \\
H_2 \\
H_3 \\
H_4 
\end{pmatrix}
= 
R_1(\beta_1)R_2(\beta_2)R_3(\beta_3)
\begin{pmatrix}
\Phi_1\\
\Phi_2\\
\Phi_3\\
\Phi_4
\end{pmatrix}
\equiv R_{\beta\alpha}\Phi_\alpha , 
\end{align}
where the orthogonal matrices
\begin{subequations}
\label{eqC4}
\begin{equation}
R_1(\beta_1)=
\begin{pmatrix}
 \cos \beta_1 & \sin \beta_1 & 0 & 0\\
-\sin \beta_1 & \cos \beta_1 & 0 & 0\\
      0     &   0        & 1 & 0\\    
      0     &   0        & 0 & 1          
\end{pmatrix},
\end{equation}
\begin{equation}
R_2(\beta_2)=
\begin{pmatrix}
      1     &   0        & 0                & 0\\          
      0     & \cos \beta_2 & \sin \beta_2   & 0 \\
      0     &-\sin \beta_2 & \cos \beta_2   & 0 \\
      0     &   0          & 0              & 1\\    
\end{pmatrix},
\end{equation}
\begin{equation}
R_3(\beta_3)=
\begin{pmatrix}
      1     &  0              &   0          & 0              \\
      0     &  1              &   0          & 0              \\ 
      0     &  0              & \cos \beta_3 & \sin \beta_3   \\
      0     &  0              &-\sin \beta_3 & \cos \beta_3   \\
\end{pmatrix},
\end{equation}
\end{subequations}
where 
$\tan \beta_1 =\frac{\sqrt{v_2^2+v_3^2+v_4^2}}{v_1}$,
$\tan \beta_2 =\frac{\sqrt{v_3^2+v_4^2}}{v_2}$ and 
$\tan \beta_3 =\frac{v_4}{v_3}$, {and $H_\beta=(H_\beta^+,(H_\beta^0+iH_\beta^{\text{odd}})/\sqrt{2})^\text{T}$.}
This basis is chosen in such a way that only the neutral component of $ H_1 $ 
acquires
a vacuum expectation value
{
\begin{align}
&\langle H_1^0 \rangle =\sqrt{v_1^2+v_2^2+v_3^2+v_4^2}\equiv v
,\hspace{0.5cm}\notag\\
&\langle H_2^0 \rangle = 0,\hspace{0.5cm}
\langle H_3^0 \rangle  = 0,\hspace{0.5cm}
\langle H_4^0 \rangle  = 0.
\end{align}}
In this way $\Phi_\alpha y_{ij}^{F\alpha}= y_{ij}^{F\alpha}R_{\alpha \beta}^T 
R_{\beta \gamma} \Phi_{\gamma}= \Y_{ij}^{F\beta} H_\beta $, and $F=U,D,N,E$;
where we have defined
\begin{align}
\label{eqC8}
\Y^{F\beta}_{ij}=R_{\beta \alpha}y_{ij}^{F\alpha}.
\end{align}
With these definitions, equation~(\ref{eq:yukawa}) becomes
\begin{align}
\mathcal{L} = -\bar{q}_L^{\prime i} H_\beta \Y_{ij}^{D\beta}d^{\prime j}_{R}
              -\bar{q}_L^{\prime i} \tilde H_\beta \Y_{ij}^{U\beta}u^{\prime j}_{R}
              -\bar{\ell}_L^{\prime i} H_\beta \Y_{ij}^{E\beta}e^{\prime j}_{R}
              -\bar{\ell}_L^{\prime i} \tilde H_\beta \Y_{ij}^{N\beta}\nu^{\prime j}_{R}+\text{h.c}.   
\end{align}
It is necessary to rotate to the fermion mass eigenstates, i.e.,
\begin{align}
f_{L,R}= U^F_{L,R}f'_{L,R},
\end{align}
where the diagonalization matrices $U_{L,R}$  are defined below, in 
section~\ref{sec:mat-diag}.
From the Lagrangian for the charged currents
\begin{align}
\mathcal{L}_{CC}=&-\frac{g}{\sqrt{2}} \bar{u}_{Li}'\gamma^{\mu} d_{Li}'W^+
                  -\frac{g}{\sqrt{2}} \bar{e}_{Li}'\gamma^{\mu} \nu_{Li}'W^-
    +\text{h.c}\notag\\
                =&-\frac{g}{\sqrt{2}} \bar{u}_{Li}\gamma^{\mu}\left(V_{_{\text{CKM}}}\right)_{ij} d_{Lj}W^+
                  -\frac{g}{\sqrt{2}} \bar{e}_{Li}\gamma^{\mu}\left(V_{_{\text{PMNS}}}\right)_{ij} \nu_{Lj}W^-
                 +\text{h.c},
\end{align}
it is possible to obtain the CKM ($V_{_{\text{CKM}}}= U^{U}_L U^{D \dagger}_L$) 
and  PMNS ($V_{_{\text{PMNS}}}= U^{E}_L U^{\nu \dagger}_L$) mixing matrices
by rotating to the fermion mass eigenstates.
In particular, we are interested in the coupling of the axial neutral current to the axion in the mass eigenstates.
\begin{align}
\mathcal{L}_{H^0} =&
              -\frac{1}{\sqrt{2}}\bar{d}_L^{\prime i}   H_\beta^0    \Y_{ij}^{D\beta}     d^{\prime j}_{R}
              -\frac{1}{\sqrt{2}}\bar{u}_L^{\prime i}   H_\beta^{0*} \Y_{ij}^{U\beta}     u^{\prime j}_{R}
              -\frac{1}{\sqrt{2}}\bar{e}_L^{\prime i}   H_\beta^0    \Y_{ij}^{E\beta}     e^{\prime j}_{R}
              -\frac{1}{\sqrt{2}}\bar{\nu}_L^{\prime i} H_\beta^{0*} \Y_{ij}^{N\beta}   \nu^{\prime j}_{R}
              +\text{h.c},\notag\\
              =&
              -\frac{1}{\sqrt{2}}\bar{d}_L^{i}   H_\beta^0    Y_{ij}^{D\beta}     d^{j}_{R}
              -\frac{1}{\sqrt{2}}\bar{u}_L^{i}   H_\beta^{0*} Y_{ij}^{U\beta}     u^{j}_{R}
              -\frac{1}{\sqrt{2}}\bar{e}_L^{i}   H_\beta^0    Y_{ij}^{E\beta}     e^{j}_{R}
              -\frac{1}{\sqrt{2}}\bar{\nu}_L^{i} H_\beta^{0*} Y_{ij}^{N\beta}   \nu^{j}_{R}
              +\text{h.c}, \notag
              \end{align}
where  $Y_{ij}^{F\beta} = \left(U^{F}_{L}   \Y^{F\beta}   U^{F\dagger}_{R}  
\right)_{ij} $.
In these expressions the mass functions in the interaction basis are:
\begin{align}\label{eq:b11}
M^{D}_{ij}= \frac{v}{\sqrt{2}} \Y_{ij}^{D1},\hspace{0.5cm}
M^{U}_{ij}= \frac{v}{\sqrt{2}} \Y_{ij}^{U1},\hspace{0.5cm}
M^{E}_{ij}= \frac{v}{\sqrt{2}} \Y_{ij}^{E1},\hspace{0.5cm}
M^{N}_{ij}= \frac{v}{\sqrt{2}} \Y_{ij}^{N1},
\end{align}
where {$v= \langle H_1^0 \rangle$} is the Higgs vacuum expectation value.

\section{Scalar potential}
\label{sec:scalars}

As studied in \cite{Giraldo:2020hwl}, the scalar sector requires four scalar doublets $\phi^\alpha$ to reproduce the mass textures of the fermion sector correctly, and two scalar singlets $S_1$ and $S_2$ that break the PQ symmetry while generating a phenomenologically viable scalar mass spectrum.  The $S_2$ singlet also gives mass to the heavy quark. The most general potential allowed by the PQ symmetry according to the charges established in Table~\ref{tab:pcontent2} is:

{\footnotesize
\begin{eqnarray}\label{eq:scalar-potential}
V(\Phi,S_i) &=& \sum_{i=1}^4\mu_{i}^{2}\Phi_{i}^{\dagger}\Phi_{i} +\sum_{k=1}^2\mu_{s_k}^2 S_k^{*}S_k+\sum_{i=1}^4\lambda_{i}\left(\Phi_{i}^{\dagger}\Phi_{i}\right)^{2}\nonumber \\
&+&  \sum_{k=1}^2\lambda_{s_k}\left(S_k^{*} S_k \right)^{2}     + \sum_{i=1}^4\sum_{k=1}^2\lambda_{is_k}\left(\Phi_{i}^{\dagger}\Phi_{i}\right)\left(S_k^{*}S_k\right)\nonumber\\
&+& \sum_{\underbrace{i,j=1}_{i<j}}^4\bigg{(}\lambda_{ij}\left(\Phi_{i}^{\dagger}\Phi_{i}\right) \left(\Phi_{j}^{\dagger}\Phi_{j}\right)+J_{ij}\left(\Phi_{i}^{\dagger}\Phi_{j}\right) \left(\Phi_{j}^{\dagger}\Phi_{i}\right)\bigg{)}
\nonumber\\
&+& \lambda_{s_1 s_2}\left(S_1^{*}S_1\right)\left(S_2^{*}S_2\right)\nonumber\\
&+&K_{1}\left(\left(\Phi_{1}^{\dagger}\Phi_{2}\right) \left(\Phi_{3}^{\dagger}\Phi_{2}\right) + h.c.\right)\nonumber\\
&+&
 K_{2}\left(\left(\Phi_{3}^{\dagger}\Phi_{4}\right) \left(\Phi_{3}^{\dagger}\Phi_{1}\right) + h.c.\right)\nonumber\\&+& F_1 \left( \left(\Phi_{2}^{\dagger}\Phi_{3}\right) S_1 +h.c.\right)\nonumber\\
&+&F_2 \left( \left(\Phi_{1}^{\dagger}\Phi_{2}\right) S_1 +h.c.\right)\nonumber\\
&+&\frac{1}{2} \left(m_{\zeta_{S_2}}\right)^2_{\text{SB}}\zeta^2_{S_2}+\frac{1}{2}\left(m_{\xi_{S_2}}\right)^2_{\text{SB}}\xi^2_{S_2}.
\end{eqnarray}}
where the terms proportional to $F_i$ are allowed by the particular choice of PQ charges and  {these couplings} $F_i$ have units of mass.  
After spontaneous symmetry breaking~(SSB), the four Higgs doublets acquire {VEVs that give mass} to all the SM particles. {The scalar doublets and singlets are written as follows:}
\begin{align}
\label{eq:higgs1}
\Phi_\alpha =& 
\begin{pmatrix}
\phi_\alpha^{+}\\
\frac{v_\alpha+h_\alpha+i\eta_\alpha}{\sqrt{2}}
\end{pmatrix},
\hspace{1cm}
\tilde\Phi_\alpha=i\sigma_2 \Phi_\alpha^{*},
\hspace{0.5cm}\notag {\alpha=1,2,3,4,}\\
S_i=&\frac{v_{_{S_i}}+\xi_{S_i}+i\zeta_{S_i}}{\sqrt{2}};\hspace{1cm} i=1,2,
\end{align}
{where the VEVs satisfy the following hierarchy:   $v_4\ll v_1, v_2\ll v_3\ll v_{S_1}\sim v_{S_2}$.}
The {scalar singlets $S_1$ and $S_2 $ break} the PQ symmetry at the high energy scale given by $v_{s_1}\approx v_{s_2}$. 
The last two terms in equation~\eqref{eq:scalar-potential} correspond to {the soft-breaking} masses of the  imaginary and the real {parts} of $S_2$,
which are generated at one  loop in the Coleman-Weinberg potential from the interaction term $\lambda_Q S_2\bar{Q}_RQ_L+\text{h.c.}$
{Additionally, we choose numerical values for the parameters of the potential~\eqref{eq:scalar-potential} in order to obtain a scalar sector mass spectrum consistent with the existing phenomenology.  The values of these parameters are:}
%
%
\begin{eqnarray}\label{eq:couplings}
\lambda_{1}&=& \lambda_{2}=\lambda_{4}=\lambda_{s_1}=\lambda_{s_2}=\lambda_{s_1 s_2}=
1,\hspace{1cm}\nonumber\\
\lambda_3&=&0.463\nonumber\\
\lambda_{ij}&=&1 \textup{ for any } i,j,  \nonumber\\
\lambda_{js_1}&=&\lambda_{js_2}=1\textup{ for any }j,\nonumber\\
J_{12}&=& J_{13}=J_{23}=J_{24}=-1,\ \  \textup{otherwise } J_{ij}=1,\nonumber\\
K_1&=&K_2=-1,\nonumber\\
F_1&=&F_2=-1 \text{GeV}.
\end{eqnarray}
In particular, the value of $\lambda_{3}$  adjusts the SM Higgs mass.
%
%
The $v_i$ are determined from the SM fermion masses and the quark mass matrix {textures, Eq.~\eqref{eq28}. }
The {VEV} $v_{s_1}$ {remains} a free parameter; {however}, this parameter is important for the axion physics due to  the {relationship}~\cite{Giannotti:2017hny},
\begin{align}
f_a=\frac{v_{s_1}}{2N}.
\end{align}
In our calculations we took $v_{s_1}\approx v_{s_2}\approx 10^{6}$GeV.
It is important to emphasize that in our model, $ f_a $ can take arbitrary values; {nevertheless}, a small $f_a$ {restricts} $\epsilon$ (Eq.~\ref{eq:parametrization}) to values close to zero.
Taking into account all these considerations, {including} Eq.~\eqref{eq:couplings}, the scalar mass spetrum~(in GeV) is:
{\footnotesize
\begin{align}
&\text{CP even}         =\{1.73\times10^6,1.\times10^6,6.54\times10^3,1.97\times10^3,\nonumber\\
&\hspace{1.8cm}1.09\times 10^3,125\},\nonumber\\
&\text{CP odd}          =\{6.54\times10^3,1.97\times10^3,1.09\times 10^3,0,0,m_{\zeta_{S_2}}\},\nonumber\\
&\text{Charged fields} = \{6.54\times10^3,1.97\times10^3,1.11\times 10^3,0\}.
\end{align}}
The mass spectrum of the scalar fields is above the TeVs scale, except {for} the SM Higgs,  which {is at} 125~GeV.
The pseudoscalar sector (CP odd fields) have two {massless} eigenstates, the axion field  and the Goldstone boson 
which is absorbed by the longitudinal component of the SM $ Z $ boson.  
A similar result is {obtained} in the charged sector, where it is possible to identify the two Goldstone bosons {required} to give mass to the SM  $W^{\pm}$ fields.

\section{diagonalization matrices}
\label{sec:mat-diag}
To compare with physical quantities, it is necessary to rotate fields 
to the mass eigenstates, i.e.,
$f_{L,R}=U^{F}_{L,R}f'_{L,R}$, where the  prime {symbol stands} for the interaction basis. 
In our formalism the {quark} mass matrices are Hermitian, {so}
the {right- and left-handed} diagonalizing matrices 
are identical; {additionally}, {we establish that the eigenvalues of the second family of quarks are negative in order to {generate} texture-zeros in some diagonal terms of the mass matrices, as indicated in}~\cite{Branco:1999nb}. {This sign is taken into account by introducing the identity matrix written as} $I_2I_2=1$
with $I_2=\text{diag}(1,-1,1)$, i.e.,
\begin{align}
M^{F}_{ij}=&\left(U^{F\dagger}\lambda^{F}U^{F}\right)_{ij}=
\left(U^{F\dagger}_L m^{F}U^{F}_R\right)_{ij}
=\frac{v}{\sqrt{2}}\mathcal{Y}^{F1}_{ij}=\frac{v}{\sqrt{2}} R_{1\alpha}y^{F\alpha}_{ij},
\end{align}
where $\mathcal{Y}_{ij}^{F\beta}$ and $R_{\alpha\beta}$ were defined 
in section~\ref{sec:mass-op}, 
$\lambda^{U,D}=\text{diag}(m_{u,d},-m_{c,s},m_{t,b})$ 
and
$m^{U,D}=\text{diag}(m_{u,d},m_{c,s},m_{t,b})$,
with similar definitions in the lepton sector, i.e.,   
$\lambda^{N,E}=\text{diag}(m_{1,e},-m_{2,\mu},m_{3,\tau})$,
$m^{N,E}=\text{diag}(m_{1,e},m_{2,\mu},m_{3,\tau})$,  and
\begin{align}\label{eq:url}
U_L^{F}= U^{F}, \hspace{1cm} U_R^{F}= I_2 U^{F},
\end{align}
where the $U^{F}$ diagonalization matrices are defined below.
  It is important to stress that the texture-zeros pattern in the matrix $\mathcal{Y}^{F1}_{ij}$   are identical to those in the original Yukawa couplings $y^{F\alpha}_{ij}$, 
  since the sum over $\alpha$ does not mix the  $i, j$ indices.
  In fact, according to equations~\eqref{eq:yij} and \eqref{eq29},  $M^{F}= \frac{v_{\alpha}}{\sqrt{2}}y^{F\alpha}_{ij}=\frac{v}{\sqrt{2}} R_{1\alpha}y^{F\alpha}_{ij}$, therefore
  $R_{1\alpha}=\frac{v_{\alpha}}{v}$. 
The diagonalization matrices are:
\begin{equation}
 U^{U\dagger}=
\begin{pmatrix}
e^{i (\phi_{C_u}+ \theta_{1u})} \sqrt{\frac{m_c m_t (A_u-m_u)}{A_u (m_c+m_u) (m_t-m_u)}} & -e^{i (\phi_{C_u}+ \theta_{2u})} \sqrt{\frac{(A_u+m_c) m_t m_u}{A_u (m_c+m_t) (m_c+m_u)}} & e^{i (\phi_{C_u}+ \theta_{3u})} \sqrt{\frac{m_c (m_t-A_u) m_u}{A_u (m_c+m_t) (m_t-m_u)}} \\
 -e^{i( \phi_{B_u}+ \theta_{1u})} \sqrt{\frac{(A_u+m_c) (m_t-A_u) m_u}{A_u (m_c+m_u) (m_t-m_u)}} & -e^{i (\phi_{B_u}+ \theta_{2u})} \sqrt{\frac{m_c (m_t-A_u) (A_u-m_u)}{A_u (m_c+m_t) (m_c+m_u)}} &e^{i (\phi_{B_u}+ \theta_{3u})} \sqrt{\frac{(A_u+m_c) m_t (A_u-m_u)}{A_u (m_c+m_t) (m_t-m_u)}} \\
 e^{i \theta_{1u}} \sqrt{\frac{m_u (A_u-m_u)}{(m_c+m_u) (m_t-m_u)}} & e^{i \theta_{2u}} \sqrt{\frac{m_c (A_u+m_c)}{(m_c+m_t) (m_c+m_u)}} & e^{i \theta_{3u}}\sqrt{\frac{m_t (m_t-A_u)}{(m_c+m_t) (m_t-m_u)}} 
\end{pmatrix},
\label{Uu}
\end{equation}
\begin{equation}
\label{diagMd}
 U^{D\dagger}=
\begin{pmatrix}
 e^{i \theta_{1d}} \sqrt{\frac{m_b (m_b-m_s) m_s}{(m_b-m_d) (m_d+m_s) (m_b+m_d-m_s)}} & -e^{i \theta_{2d}} \sqrt{\frac{m_b (m_b+m_d) m_d}{(m_d+m_s) (m_b+m_d-m_s) (m_b+m_s)}} & \sqrt{\frac{m_d (m_s-m_d) m_s}{(m_b-m_d) (m_b+m_d-m_s) (m_b+m_s)}} \\
 e^{i \theta_{1d}} \sqrt{\frac{m_d (m_b-m_s)}{(m_b-m_d) (m_d+m_s)}} & e^{i \theta_{2d}} \sqrt{\frac{(m_b+m_d) m_s}{(m_d+m_s) (m_b+m_s)}} & \sqrt{\frac{m_b (m_s-m_d)}{(m_b-m_d) (m_b+m_s)}} \\
 -e^{i \theta_{1d}} \sqrt{\frac{m_d (m_b+m_d) (m_s-m_d)}{(m_b-m_d) (m_d+m_s) (m_b+m_d-m_s)}} & -e^{i \theta_{2d}} \sqrt{\frac{(m_b-m_s) m_s (m_s-m_d)}{(m_d+m_s) (m_b+m_d-m_s) (m_b+m_s)}} & \sqrt{\frac{m_b (m_b+m_d) (m_b-m_s)}{(m_b-m_d) (m_b+m_d-m_s) (m_b+m_s)}} 
\end{pmatrix},
\end{equation}
where $\theta_{1u},\theta_{2u},\theta_{3u},\theta_{1d}$ and $\theta_{2d}$
are arbitrary phases~(a third phase for the diagonalization matrix ~\eqref{diagMd} can be absorbed by the remaining phases) that are useful for {conforming} to the $V_{\text{CKM}}=U^{U}_{L}\,U^{D\dag}_{L}$ matrix convention. 
Taking as input the SM parameters at the $Z$ pole, the best  fit values  are {given in Table~\ref{tab:eqF3}.}
\begin{table}[h]
\begin{center}
{
{\begin{tabular}{|c|c|c|c|c|c|c|}
\hline  
$\theta_{1u}$&$\theta_{2u}$&$\theta_{3u}$&$\theta_{1d}$&$\theta_{2d}$& $\phi_{C_u}$&$ \phi_{B_u}$\\
\hline
$-2.84403$     &1.85606      &$-0.00461668 $ &1.93013      &$-0.976639$    & $-1.49697  $  &0.301461\\
\hline
\hline
$A_u$      &$m_u$     &$m_c$      &$m_t$     &$m_d$      &$m_s$      &$m_b$    \\ 
\hline
1690.29~MeV&1.2684~MeV&633.197~MeV&171268~MeV&3.14751~MeV&56.1169~MeV&2910.01~MeV\\  
\hline
\end{tabular}}
\caption{{Best-fit point of the mass matrix parameters with respect to experimental data for the masses and mixing angles of the quark sector at the Z pole.}}
\label{tab:eqF3}
}
\end{center}
\end{table}

{Similarly}, in the lepton sector,
the diagonalization matrices of the mass matrices~\eqref{eq:texturalep} are:
%
\begin{align} 
 \label{32x}
      U^{N\dag}&=\begin{pmatrix}
     e^{i(\theta_{1\nu}+c_\nu)}\, 
\sqrt{\frac{m_2 m_3(A_\nu-m_1)}{A_\nu(m_2
+m_1)(m_3-m_1)}}&-e^{i(\theta_{2\nu}+c_\nu)}
\sqrt{\frac{m_1m_3(m_2+A_\nu)}{A_\nu(m_2
+m_1)(m_3+m_2)}}&e^{i(\theta_{3\nu}+c_\nu)}\,
\sqrt{\frac{m_1m_2(m_3-A_\nu)}{A_\nu(m_3
-m_1)(m_3+m_2)}}\\
&&&\\[-2mm]
e^{i\theta_{1\nu}}\sqrt{\frac{m_1(A_\nu-m_1)}{(
m_1+m_2)(m_3-m_1)}}&
e^{i\theta_{2\nu}}\sqrt{\frac{m_2(A_\nu+m_2)}{(m_2+m_1 )(m_3+m_2)}}&e^{i\theta_{3\nu}}\,
\sqrt{\frac{m_3(m_3-A_\nu)}{(m_3-m_1
)(m_3+m_2)}}\\
&&&\\[-2mm]
-e^{i(\theta_{1\nu}-b_\nu)}\sqrt{\frac{m_1(A_\nu+m_2)(m_3-A_\nu)}
{ A_\nu(m_1+m_2)(m_3-m_1)}}&
-e^{i(\theta_{2\nu}-b_\nu)}\sqrt{\frac{m_2(A_\nu-m_1)(m_3-A_\nu)}{
A_\nu(m_2+m_1)(m_3+m_2)}}&e^{i(\theta_{3\nu}-b_\nu)}\,
\sqrt{\frac{m_3(A_\nu-m_1)(A_\nu+m_2)}{A_\nu(m_3
-m_1)(m_3+m_2)}}
     \end{pmatrix}, 
       &\\ \nonumber
       \\
       \label{32}
       U^{E\dag}&=\begin{pmatrix}
     e^{i\theta_{1\ell}}\, 
\sqrt{\frac{m_\mu m_\tau(m_\tau-m_\mu)}{(m_e-m_\mu+m_\tau)(m_\mu
+m_e)(m_\tau-m_e)}}&-e^{i\theta_{2\ell}}
\sqrt{\frac{m_em_\tau(m_e+m_\tau)}{(m_e-m_\mu+m_\tau)(m_\mu
+m_e)(m_\tau+m_\mu)}}&
\sqrt{\frac{m_em_\mu(m_\mu-m_e)}{(m_e-m_\mu+m_\tau)(m_\tau
-m_e)(m_\tau+m_\mu)}}\\
&&&\\[-2mm]
e^{i\theta_{1\ell}}\sqrt{\frac{m_e(m_\tau-m_\mu)}{(m_\mu
+m_e)(m_\tau-m_e)}}&
e^{i\theta_{2\ell}}\sqrt{\frac{m_\mu(m_e+m_\tau)}{(m_\mu+m_e )(m_\tau+m_\mu)}}&
\sqrt{\frac{m_\tau(m_\mu-m_e)}{(m_\tau-m_e
)(m_\tau+m_\mu)}}\\
&&&\\[-2mm]
-e^{i\theta_{1\ell}}\sqrt{\frac{m_e(m_e+m_\tau)(m_\mu-m_e)}
{(m_e-m_\mu+m_\tau)(m_\mu+m_e)(m_\tau-m_e)}}&
-e^{i\theta_{2\ell}}\sqrt{\frac{m_\mu(m_\tau-m_\mu)(m_\mu-m_e)}{
(m_e-m_\mu+m_\tau)(m_\mu+m_e)(m_\tau+m_\mu)}}&
\sqrt{\frac{m_\tau(m_\tau-m_\mu)(m_e+m_\tau)}{(m_e-m_\mu+m_\tau)(m_\tau
-m_e)(m_\tau+m_\mu)}}
     \end{pmatrix},
\end{align}
%
where $\theta_{1\ell}$, $\theta_{2\ell}$, $\theta_{1\nu}$, $\theta_{2\nu}$, $\theta_{3\nu}$ 
are necessary phases in order to adjust to the established convention for the PMNS mixing matrix~\cite{Esteban:2018azc}~\footnote{
NuFIT collaboration~(http://www.nu-fit.org/?q=node/211)(with SK atmospheric 
data).};  and  $c_\nu$ and $b_\nu$  are the phases of $C_\nu$ and  $B_\nu$ in the neutral mass matrix $M^N$ in Eq.~\eqref{eq:texturalep}.  The best fit values for these quantities are shown in Table~\ref{tab:input}. 
\begin{center}
\begin{table}[h]
 \begin{tabular}{|c|c|c|c|c|c|c|}
 \hline
  $\theta_{1\ell}$& $\theta_{2\ell}$&$\theta_{1\nu}$&$\theta_{2\nu}$&$\theta_{3\nu}$&$c_\nu$&$b_\nu$\\[1mm]\hline
&&&&&&\\[-2mm]
0.154895 &2.01797& $-0.835504$  & 2.21169
&1.81786&1.01608&2.03726\\[1mm]
\hline\hline
&&&&&&\\[-3mm]
$A_\nu\,(\text{eV})$&$m_e\,(\text{MeV})$&$m_\mu\,(\text{MeV})$&$m_\tau\,(\text{MeV}
)$&$m_1\,(\text{eV})$&$m_2\,(\text{eV})$&$m_3\,(\text{eV})$\\[1mm]\hline
&&&&&&\\[-2mm]
0.0251821&0.5109989461&105.6583745 &1776.86&0.00353647&0.00929552&0.0504034\\[1mm]\hline
\end{tabular}
\caption{Best fit values.}
\label{tab:input}
\end{table}
\end{center}

 \section{Axion decay into photons}
 \label{sec:appendix-gamma}
 In the SM, $B^{\mu}=\cos\theta_W A^{\mu}-\sin\theta_W Z^{\mu}$ and $W^{3\mu}=\sin\theta_W A^{\mu}+\cos\theta_W Z^{\mu}$,
 where $A^{\mu}$ and $Z^{\mu}$ are the SM fields for the photon and  $Z$ gauge bosons,  replacing 
 these expresions in Eq.~\eqref{eq:gauge}  we obtain
 \begin{align}
 \mathcal{L}&\supset
 -c_1^{\text{eff}}\frac{\alpha_1}{8\pi}
 \frac{ a}{\Lambda_{\text{PQ}}}B_{\mu\nu}\tilde B^{\mu\nu}
 -c_2^{\text{eff}}\frac{\alpha_2}{8\pi}
 \frac{ a}{\Lambda_{\text{PQ}}}W_{\mu\nu}^3\tilde W^{3\mu\nu}
 \notag\\
 &=-\frac{\alpha}{8\pi}
 (c_1^{\text{eff}}+c_2^{\text{eff}})
 \frac{a}{\Lambda_{\text{PQ}}}
 F_{\mu\nu}\tilde F^{\mu\nu}
 \notag\\
 &-\frac{\alpha}{8\pi  c_W^2 s_W^2}
 (s^4_W c_1^{\text{eff}}+c^4_W c_2^{\text{eff}})\frac{a}{f_a} Z_{\mu\nu}\tilde Z^{\mu\nu}\notag\\
 &-\frac{2\alpha}{8\pi  c_W s_W }
 (c^2_W c_2^{\text{eff}}-c^2_W c_1^{\text{eff}})
 \frac{a}{\Lambda_{\text{PQ}}}
 F_{\mu\nu}\tilde Z^{\mu\nu}\notag\\
 &=e^2C_{\gamma\gamma}
 \frac{a}{\Lambda_{\text{PQ}}}
 F_{\mu\nu}\tilde F^{\mu\nu}
 +\frac{e^2 C_{ZZ}}{  c_W^2 s_W^2}
 \frac{a}{\Lambda_{\text{PQ}}} 
 Z_{\mu\nu}\tilde Z^{\mu\nu}
 +\frac{2e^2C_{\gamma Z}}{  c_W s_W }
 \frac{a}{\Lambda_{\text{PQ}}}
 F_{\mu\nu}\tilde Z^{\mu\nu}
 \end{align}
 \begin{align}
 c_1^{\text{eff}}& = c_1-\frac{1}{3}\Sigma q+\frac{8}{3}\Sigma u+\frac{2}{3}\Sigma d
 -\Sigma l+ 2\Sigma e\\
 c_2^{\text{eff}}& = c_2-3\Sigma q-\Sigma l
 \end{align}
 where $\Sigma f\equiv f_1+f_2+f_3$ is the sum of the PQ charges of the three families. 
 There are similar definitions for the interaction of the axion with the gluons
 \begin{align}
  -c_3^{\text{eff}}\frac{\alpha_3}{8\pi}
 \frac{ a}{\Lambda_{\text{PQ}}}G_{\mu\nu}^a\tilde G^{a\mu\nu}
 =  g_s^2C_{GG}
 \frac{ a}{\Lambda_{\text{PQ}}}G_{\mu\nu}^a\tilde G^{a\mu\nu},
 \end{align}
 where $c_3^{\text{eff}}=c_3-2\Sigma q+\Sigma u +\Sigma d -A_{Q}$,
 in our particular case $c_i=0$.
 In axion phenomenology, it is usual to define
 \begin{align}
  C_{\gamma\gamma}
  =& -\frac{1}{32\pi^2}
  (c_1^{\text{eff}}+c_2^{\text{eff}}),
  \hspace{1.4cm}
  C_{ZZ}
  = -\frac{1}{32\pi^2}
  (s_W^4 c_1^{\text{eff}}+c_W^4 c_2^{\text{eff}}),
  \notag\\
   C_{\gamma Z}
  =& -\frac{1}{32\pi^2}
  (c_W^2 c_2^{\text{eff}}-c_W^2 c_1^{\text{eff}}),
 \hspace{0.5cm}
 C_{GG}= -\frac{1}{32\pi^2}c_3^{\text{eff}}.
 \end{align}
 The decay widths of an axion decaying  in two photons and a $Z$ decaying in an axion and a photon are~\cite{Bauer:2017ris}
 \begin{align}
 \Gamma(a\rightarrow \gamma\gamma)
 = &\frac{4\pi \alpha^2 m_a^3}{\Lambda^2_{\text{PQ}}}
 \lvert C_{\gamma\gamma}^{\text{eff}}\rvert^2,\notag\\
 \Gamma(Z\rightarrow \gamma a)
 = &\frac{8\pi \alpha(m_Z) m_Z^3}{3s_W^2c_W^2\Lambda^2_{\text{PQ}}}
 \lvert C_{\gamma Z}^{\text{eff}}\rvert^2 
 \left(1-\frac{m_a^2}{m_Z^2}\right)^3.
 \end{align}
 Another possible decay channel of the axion in two photons is due to the mixing between the axion and the pion since the latter can decay in two photons, this decay mode generates an additional correction that only depends on the couplings of the axion to the gluons~\cite{Alonso-Alvarez:2018irt}
 \begin{align}
  C_{\gamma\gamma}^{\text{eff}}
  =& -\frac{c_3^{\text{eff}}}{32\pi^2}
  \left(\frac{c_1^{\text{eff}}+c_2^{\text{eff}}}{c_3^{\text{eff}}}-2.03\right),\notag\\
   C_{\gamma Z}^{\text{eff}}
  =& -\frac{c_3^{\text{eff}}}{32\pi^2}
  \left(\frac{c_W^2 c_2^{\text{eff}}-c_W^2 c_1^{\text{eff}}}{c_3^{\text{eff}}}-0.74/2\right).
  \end{align}
  It is usual to define $\Lambda_{\text{PQ}} =\lvert c_3^{\text{eff}} \rvert f_a$.
  
 \begin{align}
\frac{E}{N}=  \frac{c_1^{\text{eff}}+c_2^{\text{eff}}}{c_3^{\text{eff}}}.
 \end{align} 
The axion-photon interaction is given by
\begin{align}
g_{a\gamma\gamma}= \frac{4e^2 C_{\gamma\gamma}^{\text{eff}}}{\Lambda_{PQ}}
=  -\frac{\alpha} {2\pi f_{PQ}}\left(\frac{E}{N}-2.03\right)
\end{align}
where $\alpha=\frac{e^2}{4\pi}$.
 Due to the gluon-axion interaction, the axion  gets a mass term, which is described  at low energies  as  an axion-pion interaction~\cite{diCortona:2015ldu}
 \begin{align}
 m_a= 5.7(7)\mu eV \left(\frac{10^{12} \text{GeV}}{f_a}\right).
 \end{align}
\bibliographystyle{ieeetr}
\bibliography{biblio3,BiblioNoInspire}

\end{document}